\documentclass[apjl]{emulateapj}
\usepackage{amsmath}
\usepackage{multirow}
\usepackage[normalem]{ulem}
\usepackage{booktabs}

\shorttitle{DIBs at different physical conditions}
\shortauthors{Kos et al.}

\begin{document}

\title{Properties of Diffuse Interstellar Bands at Different Physical Conditions of the ISM}

\author{J. Kos\altaffilmark{1}, T. Zwitter\altaffilmark{1,2}}
\affil{$^1$Faculty of Mathematics and Physics, University of Ljubljana, Jadranska 19, 1000 Ljubljana, Slovenia; \\ e-mail: janez.kos@fmf.uni-lj.si}
\affil{$^2$Center of Excellence SPACE-SI.A\v{s}ker\v{c}eva cesta 12, 1000 Ljubljana, Slovenia}

\begin{abstract}
Diffuse interstellar bands (DIBs) can trace different conditions of the ISM along the sightline toward the observed stars. A small survey was made in optical wavelengths, producing high resolution and high signal to noise spectra.
We present measurements of 19 DIBs' properties in 50 sightlines towards hot stars, distributed at a variety of galactic coordinates and interstellar reddening. Equivalent widths were obtained by fitting asymmetric Gaussian and variable continuum to DIBs. Conditions of the ISM were calculated from 8 atomic and molecular interstellar lines. Two distinctively different types of DIBs were identified, by carefully comparing correlation coefficients between DIBs and reddening and by different behaviour in UV shielded ($\zeta$) and non-shielded ($\sigma$) sightlines. A ratio of DIBs at 5780~\AA\ and~5797 \AA\ proved to be reliable enough to distinguish between two different sightline types. Based on linear relations between DIB equivalent width and reddening for $\sigma$ and $\zeta$ sightlines, we divide DIBs into type {\sc i} (where both linear relations are similar) and type {\sc ii} (where they are significantly different). Linear relation for $\zeta$ type sightlines always show a higher slope and larger x-intercept parameter than the relation for $\sigma$ sightlines. Scatter around the linear relation is reduced after the separation, but it does not vanish completely. This means that UV shielding is the dominant factor of the DIB equivalent width vs.\ reddening relation shape for $\zeta$ sightlines, but in $\sigma$ sightlines other physical parameters play a major role. No similar dependency on gas density, electron density or turbulence was observed. A catalog of all observed interstellar lines is made public.
\end{abstract}


\section{Introduction}
Diffuse interstellar bands (DIBs) are absorption lines at the visual and near visual wavelengths observed in spectra of reddened stars \citep{herbig95}. Their carriers are not known and so the identification of the DIBs remains one of the longest-standing problems in the astronomical spectroscopy. Several carriers from the family of fulerenne molecules and polycyclic aromatic hidrogencarbonates were proposed \citep[e.g.][]{leidlmair11, salama99, zhou06}, but only some approximate wavelength matches were observed. Naphthalene and anthracene matched DIB 5450 \citep{iglesias08,iglesias10} in some sightlines, diacetylene matched profile of DIB 5069 \citep{krelowski10} and linear C$_3$H$_2$ matched DIBs 5450 and 4881 \citep{maier11}. However this can as well be a coincidence due to a large number of molecular lines, DIBs and interfering stellar lines \citep{gala11}. 

Studies of DIBs are mostly focused into characterization of the high resolution profiles of the DIBs \citep[e.g.][]{krelowski97,gala02,kerr96} and sampling of many different sightlines within the Galaxy \citep[e.g.][]{vos11,friedman11,herbig93,krelowski99,thorburn03,weselak08,snow77}. Most studies either deal with only the strongest DIBs and  a large number of sightlines or high SNR spectra with a lot of observed DIBs but a small sample of different sightlines. Only more recent surveys accumulate a large number of spectra with good SNR. The goal of our study is to detect weaker DIBs in a statistically representative number of spectra -- this puts our survey in between the two extreme cases -- and study details in the correlation between DIBs, reddening and conditions of the ISM.

Two sightline categories have been described \citep{krelowski92,krelowski94} and named $\zeta$ and $\sigma$ that correspond to UV shielded and not shielded sightlines, probing cloud cores and external regions respectively. \citet{vos11} shows that there are fundamental differences between the two groups in correlations between DIBs, reddening and gas.

Around 400 DIBs are known \citep{hobbs09}, but only a minority of these is strong enough to be useful for a more detailed study. \citet{hobbs08} and \citet{jenniskens94}(hereafter JD94) give good catalogues of DIBs, the later one being used in this paper to identify DIBs in our spectra.

In this paper we investigate relations between 19 observed DIBs, atomic lines of Na {\sc i}, Ca {\sc i}, Ca {\sc ii}, molecular lines of CH, CH$^+$ and CN and reddening in a sample of 50 different sightlines toward O and early B type stars. 
Section \ref{sec:obs} presents the spectroscopic data used in this paper. In section \ref{sec:asim} we propose in this field previously undescribed method of fitting DIBs with asymmetric Gaussian functions, giving better results for the equivalent width than a simple flux integration method.  In section \ref{sec:common} we investigate sorting DIBs into different families. Sections \ref{sec:r1} and \ref{sec:r2} present measured conditions of the ISM and their impact on correlations between DIBs and reddening. The last section contains some discussion and conclusions.

\section{Observations and data reduction}
\label{sec:obs}
\subsection{Observation}
Spectra used in this paper were obtained with the 1.82 meter telescope of the Observatory of Padua in Asiago and an echelle spectrograph. Observations were made in 6 observing sessions from January 16th 2011 to November 26th 2012. A setup used for this study covered the spectral range from 3700 \AA\ to 7300 \AA\ with a resolution power around 23,000 in 34 echelle orders. Exposure times were sufficient to achieve a signal to noise ratio above 300 for most of the spectral range in our interest. Peak {S$/$N} ratio is around 5700 \AA\ and exceeds 500 in majority of the spectra. Exception are some of the spectra from the first observing session that have a somewhat lower {S$/$N}. {S$/$N} drops to 65\% of peak {S$/$N} around 4300 \AA\ and 6700\AA\ and below 20\% at 3900 \AA .

\begin{figure}[!ht]
\centering
\includegraphics[width=\columnwidth]{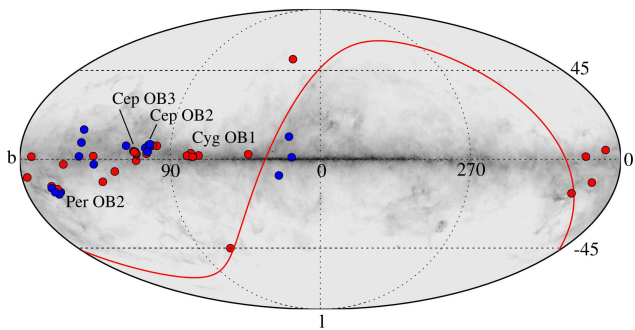}
\caption{All-sky map of all observed stars. White (blue) are $\zeta$ and black (red) are $\sigma$ sightlines, as defined in section \ref{sec:r2}. Thick gray (red) line is the celestial equator. Underlying image shows galactic dust \citep{schlegel98}. Color version of this figure is available only in the electronic edition.}
\label{fig:allsky}
\end{figure}

Program stars were selected with a prior knowledge of reddening and spectral type only. Spectral types are limited to B3 and hotter, except for one A2 star and two B5 stars. A2 star has a high rotational velocity and should not corrupt the ISM spectrum. 50 sightlines cover different parts of the Galactic plane that were observable. Some of the stars belong to the same OB associations, as marked in figure \ref{fig:allsky}. We tried to select stars in a way that the reddening range was covered as uniformly as possible.

\begin{figure}[!ht]
\centering
\includegraphics[width=\columnwidth ,height=5.6cm]{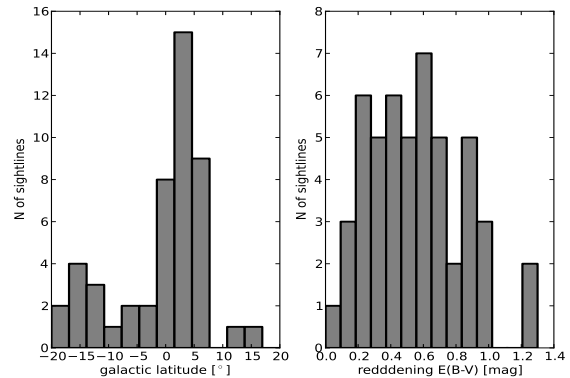}
\caption{Left: Distribution of sightlines on galactic latitude. Two sightlines at $b=51^\circ$ and $b=-44^\circ$ are excluded from the histogram. Right: Distribution of sightlines on reddening.}
\label{fig:redd_hist}
\end{figure}

Only stars with color excess up to 1.2 magnitude were observed due to moderate light collecting power of the telescope (stars brighter than 8.5 mag in V band can be observed with reasonable exposure times). Histograms in figure \ref{fig:redd_hist} show distribution of sightlines on galactic latitude and reddening. Table \ref{tab:list} shows basic parameters of individual observed stars. 

\subsection{Data reduction}

Data reduction was done with NOAO's Image Reduction and Analysis Facility (IRAF). Flat field frames, dark frames and bias frames were taken every observing session. Spectra were normalized, but no atmospheric extinction correction was applied, because all of the observed DIBs lie at wavelengths with no or negligible telluric lines. There is only one major DIB at 6281 \AA\ (JD94), that is corupted by telluric lines. We left this one out of our study. Spectra were shifted into Heliocentric velocity frame. 

\begin{figure*}[!ht]
\centering
\includegraphics[width=\textwidth]{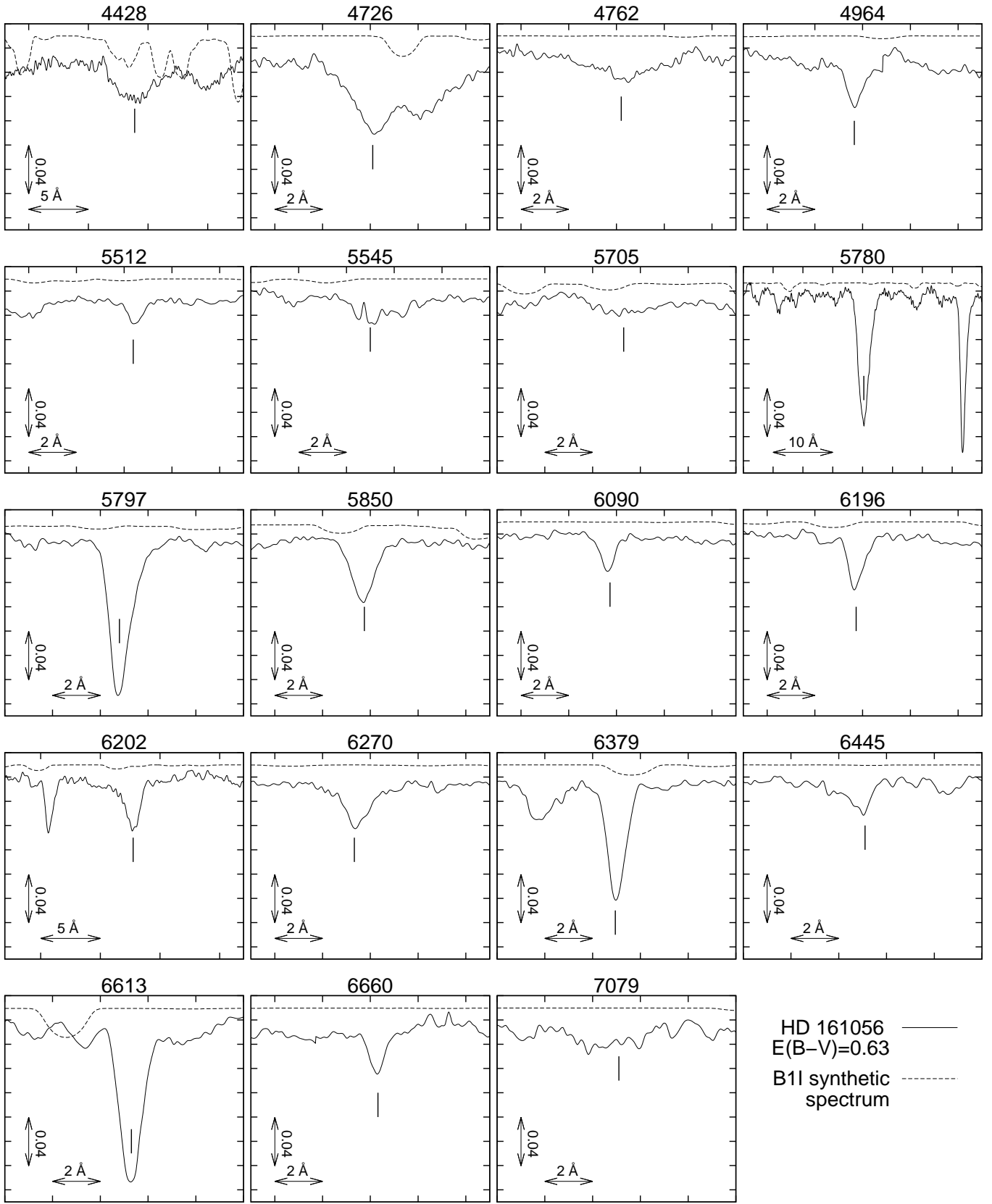}
\caption{All observed DIBs in a spectrum of star HD 161056 (solid line) and a synthetic spectrum of a B1 giant with Solar metallicity and rotational velocity of 50 km s$^{-1}$ (dashed line). Synthetic spectrum is shifted in flux for 0.03. All spectra are normalized. Each box is 10~\AA\ wide and 0.17 units high, except the boxes for DIBs at 4428~\AA\ and at 6202~\AA , which is 20~\AA\ wide and box for DIB at 5780~\AA , that is 40~\AA\ wide. Box is centered at the wavelength indicated on the top. This is also the name used for each DIB throughout the paper. DIBs of interest are indicated with a vertical dash.}
\label{fig:all}
\end{figure*}

Figure \ref{fig:all} shows all 19 DIBs that were measured in all of the spectra. A comparison with a B1 giant spectrum is also shown. Many more DIBs can be detected, especially in highly reddened stars, but we only focused on those, visible in all 50 sightlines. Stellar spectrum was not subtracted from the observed spectrum due to simplicity and because most stellar spectra lack any lines at DIBs' positions.

\section{Measurements}
\label{sec:asim}

\subsection{Fitting asymmetric Gaussians}

Most of the DIBs are not corrupted by stellar lines. Exceptions are DIBs at 4428 \AA , where the DIB is blended with several stellar lines and the combined spectrum is too complicated for deblending, and DIBs at 4726 \AA\ and 5705 \AA , which are blended with one stellar line at 4727.3 \AA\ and 5701.2\AA\ respectively. These lines can be deblended in some of the cases. 5705 seems to suffer much less due to blending than other two lines and is therefore the only one of the above 3 DIBs included in the further analysis. The other complicated cases are DIBs 5780 and 6202 which are blended with other wider DIBs. In the case of 5780 the blended DIB is 7.5 times wider (JD94) and much shallower and was treated as a part of the local continuum. DIB 6202 is blended with two other DIBs, one with a FWHM of 1.2 \AA\ at 6203.19 \AA, and the other one with FWHM of 11.7 at 6207.8 \AA\ (JD94). DIBs 6202 and the blended one at 6203.19 \AA\ were considered as a single DIB, because we were unable to deblend them. From figure \ref{fig:all} it can be seen that the two DIBs are not separated and are represented well with a single asymmetric Gaussian, a function form, we use also for other DIBs (see below). DIB 6202 is never contaminated by a C {\sc iii} stellar line at 6205.6 \AA. We note that the measured profile of 7079 is notably different from the one reported in the literature. At the wavelength of a narrow DIB we detect a depletion of the continuum, 3 times as wide as the DIB 7079 in JD94. We checked for possible misfitted continuum due to telluric lines of H$_2$O that lie in this region, but could not find any reason for it. We measured the profile of the depletion, excluding few narrow and weak telluric lines that blend with the depletion when fitting the profile. Precise enough removal of the telluric contamination was not possible due to the variable atmospheric conditions in most of the nights. Due to the discrepancy with the literature and possible strong telluric contamination we excluded this DIB from the further analysis. Other DIBs can be extracted by subtracting a simple local continuum fit. From here on, only 16 DIBs will be studied, excluding 4428, 4726 and 7079 from the original 19 measured DIBs.

Image \ref{fig:all} shows all DIBs observed and discussed in this paper. At least for strongest DIBs it is evident, that their profile cannot be represented by a single Gaussian curve. The deviation from the Gaussian curve is not due to multiple components with different radial velocities, but due to unresolved fine structure of the DIB itself. The next order of approximation after a simple Gaussian curve is to allow for the profile to be asymmetric. Asymmetry can be introduced into the Gaussian curve in many ways, two of them being presented in this paper. Both gave satisfactory results when representing the profile of DIBs.

Asymmetric Gaussian is a curve composed from two Gaussians with different widths. We choose the following definition:
\begin{equation}
\begin{array}{l}
\mathcal{A}(x; \mu, \sigma_1, \sigma_2, C)=C \frac{1}{\sqrt{2\pi} \sigma}\left[ (1-\chi(x-\mu))\exp{\frac{-(x-\mu)^2}{2\sigma_1^2}}\right.\\ \qquad\qquad\qquad\qquad\qquad\qquad\qquad \left.+\chi(x-\mu)\exp{\frac{-(x-\mu)^2}{2\sigma_2^2}} \right] \\[0.5cm]

\sigma=\frac{\sigma_1+\sigma_2}{2} \\[0.5cm]

\chi(x)=\left\{ 
\begin{array}{rl}
	1 & \mbox{if $x>0$},\\
	0 & \mbox{otherwise.}
\end{array} \right.

\end{array}
\label{eq:asimg}
\end{equation}

$\sigma_1$ and $\sigma_2$ are widths of both Gaussian components, $\mu$ is the position of the Gaussian along the $x$ axis. It corresponds to the position of the maximum or minimum of the spectral line. $C$ is the area under the curve and represents the equivalent width of a spectral line in a normalized spectrum. Such function preserves most of the properties of the usual Gaussian function. Area under the curve will always have the value of $C$ and it is continuously differentiable at $(x-\mu)=0$. Asymmetry of the spectral line can be expressed with an asymmetry index
\begin{equation}
a=\frac{\sigma_1-\sigma_2}{\sigma_1+\sigma_2}.
\end{equation}
where a negative value indicates the line profile is wider toward the red part of the spectrum and a positive value means a widening toward the blue part of the spectrum.

\begin{figure}[!ht]
\centering
\includegraphics[width=0.95\columnwidth]{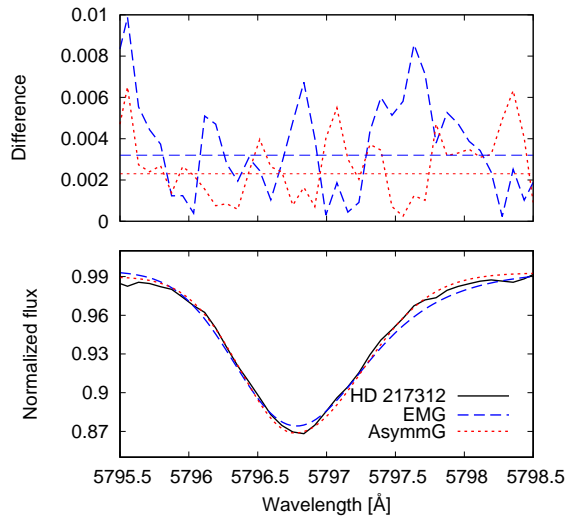}
\caption{Comparison of a fit to DIB 5797 (solid curve). Dashed curve is exponentially modified Gaussian (EMG), dotted curve is asymmetric Gaussian (AsymmG). Top plot shows differences between spectrum and both fitted functions. Horizontal lines mark average difference in the plotted range. A linear local continuum was subtracted before the fitting.}
\label{fig:emgfit}
\end{figure}

Exponentially modified Gaussian is a convolution between a Gaussian and an exponential function. It can be analytically expressed as
\begin{equation}
\begin{array}{l}
\mathcal{M}(x; \mu, \sigma, \lambda, C)= C\left[ \frac{\lambda}{2} \exp\left( \frac{\lambda}{2} \left( 2\mu + \lambda \sigma^2 -2x \right) \right)\right.\\ \qquad\qquad\qquad\qquad\qquad\qquad\qquad \left. \mathrm{erfc}\left( \frac{\mu + \lambda \sigma^2 -x}{\sqrt{2}\sigma} \right) \right] \\[0.5cm]
\mathrm{erfc}(x)=1-\mathrm{erf}(x).
\end{array}
\label{eq:emg}
\end{equation}
$\sigma$ is width of a Gaussian part, $\lambda$ is a parameter of an exponential part, $\mu$ is the mean value of the Gaussian and $C$ is the area under the curve. In this case $\mu$ does not correspond to the maximum or minimum of the function and the minimum must be calculated numerically. However the first three moments of the exponentially modified Gaussian are easy to express as:
\begin{equation}
\begin{array}{r l}
\mathrm{mean}=&\mu+\frac{1}{\lambda}\\
\mathrm{variance}=&\sigma^2+\frac{1}{\lambda^2}\\
\mathrm{skewness}=&\frac{2}{\sigma^3\lambda^3}\left( 1+\frac{1}{\sigma^2\lambda^2}\right)^{-3/2}
\end{array}
\label{eq:moments}
\end{equation}

After comparing fits of both functions (figure \ref{fig:emgfit} shows an example of fits for one DIB) to some of the strong DIBs in different spectra, we decided to use asymmetric Gaussian only. Fits with asymmetric Gaussian were slightly better and the Levenberg-Marquardt minimization of 4 parameters proved to be much more computationally stable with asymmetric Gaussian than exponentially modified Gaussian. Minimization of  exponentially modified Gaussian was even less stable when introducing linear or quadratic continuum.  Simpler evaluation of basic 4 parameters also prevailed toward the use of asymmetric Gaussian for further analysis.

A fit of two Gaussians was considered as well, but with 6 parameters in addition to 2 or 3 parameters of the continuum, the fit was badly constrained.

\begin{figure}[!ht]
\centering
\includegraphics[width=\columnwidth]{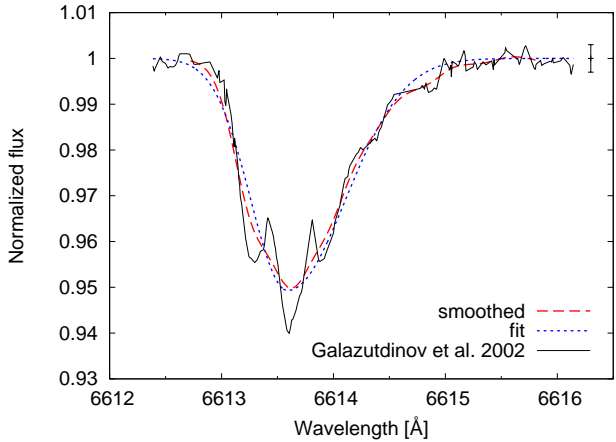}
\caption{Smoothed spectrum of DIB 6613 from \citet{gala02}, fig. 5 and asymmetric Gaussian fit. Resolution of the smoothed spectrum corresponds to the resolution of our spectra. Error bar marks typical noise of our spectra.}
\label{fig:smooth}
\end{figure}

Initial parameters were estimated interactively by marking limits, center of each DIB and area where continuum is fitted. The initial equivalent width is calculated by integrating the spectrum between both limits. Other initial parameters, including linear local continuum, are estimated from manually marked points on the spectrum. DIB, together with the continuum, is then fitted by Levenberg-Marquardt method. Errors of the final parameters are calculated from a covariance matrix. All the lines were fitted in one sitting. All fits were visually inspected to assure that the Levenberg-Marquardt method converged toward expected values.

To be sure that a modified Gaussian correctly summarizes all the information in the observed DIB profile, we took an image of the DIB 6613 from \citet{gala02}, traced the DIB profile and smoothed to the resolution of our spectra. DIB 6613 shows the most prominent substructure and should be a good test of the performance of the asymmetric Gaussian fit. Figure \ref{fig:smooth} shows the result. Differences between the asymmetric Gaussian fit and smoothed spectrum are well below the noise level of our spectra. 

\subsection{Comparison of fitted equivalent width with integrated equivalent width}
Equivalent width is defined by integration of a spectrum over a region of interest: 

\begin{equation}
W=\int_{\lambda_1}^{\lambda_2}\left( \frac{F_c-F_\lambda}{F_c}\right)d\lambda,
\end{equation}
where $\lambda_1$ and $\lambda_2$ are integration limits on each side of the line and $F_c$ and $F_\lambda$ are fluxes of continuum and observed spectrum. With $F_\lambda$ approximated with a suitable smooth function one can reduce the number of free parameters and so the error of the derived equivalent width.

Another possibility is averaging some spectra with strong DIBs and good {S$/$N} in order to get an average shape of DIBs. This profile is than fitted to other spectra with only 2 parameters: position and strength \citep{chen12,raimond12}. This method is sensitive to systematic errors and noise manifestation in an averaged DIB profile for weak bands. 

We have to keep in mind that a large portion of error comes from insufficient knowledge about the continuum. Fitting some profile to a spectral line will not reduce this part of the error. We will show, that the error of the continuum determination is smaller than the error of the integrated equivalent width, but equal or larger than the error of the fitted profile. Therefore the representation of the observed line profile with a smooth function significantly reduces the combined errors in {S$/$N} $=300$ to {S$/$N} $=400$ spectra.

All DIBs were fitted with an asymmetric Gaussian, even if their profile appears symmetric at visual inspection. Other atomic and molecular lines were fitted with an ordinary symmetric Gaussian profile, except the sodium lines where the equivalent widths were integrated instead of fitted.

\begin{figure*}[!ht]
\centering
\includegraphics[width=\textwidth]{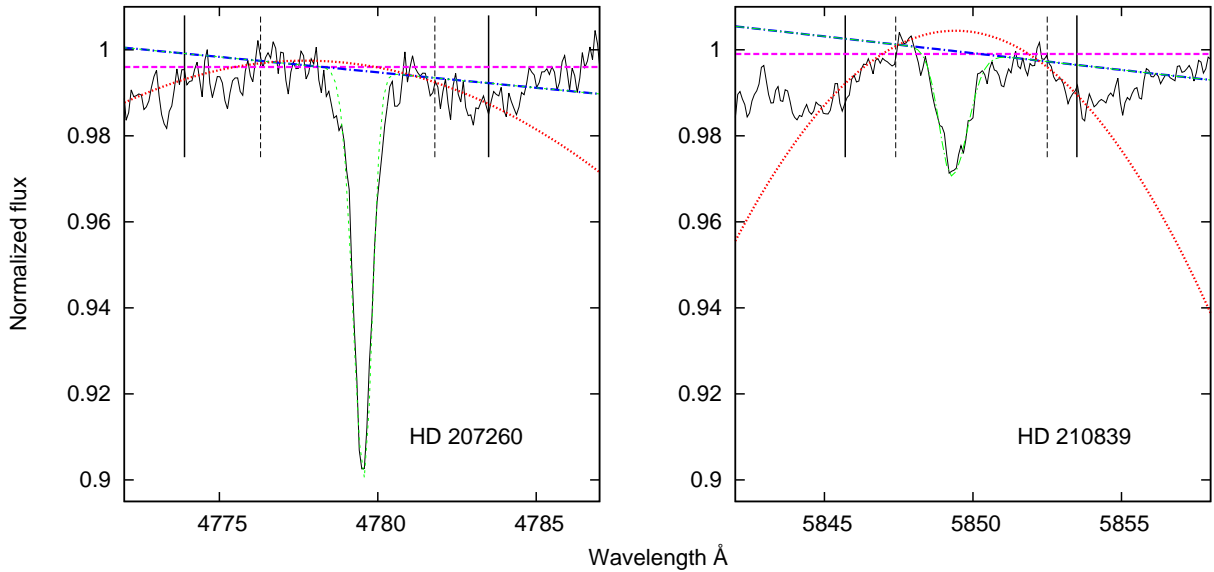}
\caption{Example showing different continuum fits. On the left is an average case (fairly strong line with simple but not linear local continuum) and on the right is the worst case we encountered (weak line with complicated local continuum). Lines plotted are: spectrum (full black line), constant fit to continuum (dashed magenta), linear fit to continuum (dash-dotted blue), quadratic fit to the continuum (dotted red) and fit to DIB with linear continuum (thin dashed green). Full vertical lines mark the boundaries of the quadratic continuum fit and dashed vertical lines mark the boundaries within the two types of linear continua were fitted. In the left case, differences in equivalent width between linear vs.\ constant and linear vs.\ quadratic fit are 2 m\AA\ and 5 m\AA\ respectively and in the right case differences are 6 m\AA\ and 21 m\AA . Differences due to three different continuum fits are calculated within dash lined boundaries. Asymmetric Gaussian for DIB and its continuum shape are fitted together as a function with 5 (constant continuum), 6 (linear continuum) or 7 (quadratic continuum) parameters.}
\label{fig:cont}
\end{figure*}

By fitting the local continuum by polynomials of different orders, we get different continuum estimates. If the spectrum around DIB is filled with other spectral lines or uneven continuum, different polynomials will give different continuum fits. If the continuum strongly depends on the type of polynomial used for the area of the DIB, a larger error is assigned to this DIB's equivalent width.

Figure \ref{fig:cont} shows how the error in continuum determination influences equivalent width measurement. We used linear fit in most of the cases in the final equivalent width measurements. A constant continuum has never been forced. In some cases (around 30\%) where the continuum appeared non-linear a quadratic fit was used. Errors in continuum determination are hard to calculate precisely. Our estimated errors on continuum determination alone vary from around 2 to 12 m\AA\ for narrow and wide DIBs respectively.

\begin{figure}[!ht]
\centering
\includegraphics[width=\columnwidth]{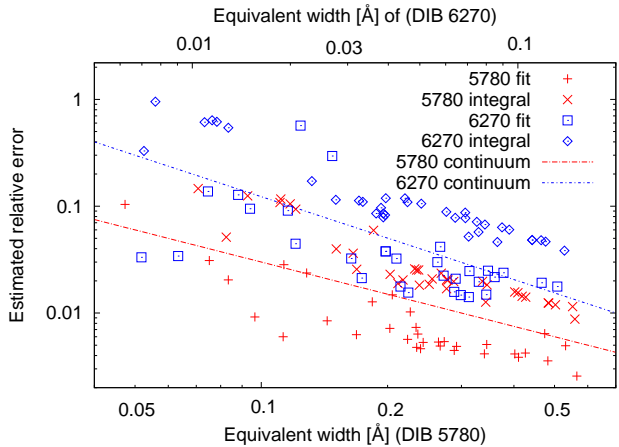}
\caption{Relative error dependence on equivalent width calculated by fitting asymmetric Gaussian (pluses and squares) and integrating the spectrum (crosses and diamonds). Errors for one strong (5780 -- dashed line) and one weak (6270 -- dotted line) DIB are shown. Errors for integrated equivalent width were calculated by monte carlo method. Two lines show estimated contribution to the relative error by a continuum fit alone. Plot shows that errors of integrated equivalent width are about two times larger than errors of the continuum. Errors of the fitted profiles are smaller than those of the continuum. Color version of this figure is available only in the electronic edition.}
\label{fig:errors}
\end{figure}

This error is larger than the error of equivalent width calculated from  the fitted profiles, but about two times smaller than the error of equivalent widths obtained by a simple sum of pixel fluxes over the DIB's profiles as shown in figure \ref{fig:errors}. By fitting DIBs by asymmetric Gaussians, the error is driven by insufficient knowledge about the continuum, not by the noise in the DIB profile. This justifies the use of asymmetric Gaussians as DIBs' profile templates.

Fitting asymmetric Gaussians or any other simple asymmetric functions can not be extrapolated to high {S$/$N} or high resolution spectra. Some DIBs have very well known substructure \citep{gala02,sarre95} that is significantly different from asymmetric shapes that are assumed in this work. Resolution and {S$/$N} of our spectra is too low to detect this substructure. Even the strongest DIBs in our spectra with known substructure only show asymmetry and no substructure. If the substructure was detectable, systematic errors would arise and the statements above would not hold any more.

\subsection{Comparison with the literature}
\label{sec:comp}
\begin{figure}[!ht]
\centering
\includegraphics[width=\columnwidth]{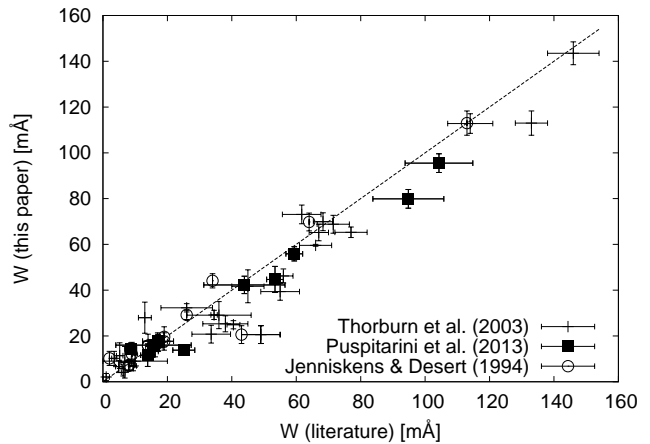}
\caption{Equivalent widths of DIBs from table \ref{tab:match} plotted as our measurements vs.\ the measurements from the literature. 1:1 relation is plotted as a dashed line.}
\label{fig:match}
\end{figure}

To verify the accuracy and consistency of our measurements with published observations, we compare our equivalent widths of 16 DIBs with those available in the literature \citep{thorburn03, jenniskens94, lucky13}. Here we used the works that include most of our DIBs. Table \ref{tab:match} shows the comparison for DIBs measured in the spectra of 4 stars discussed in the literature. There are major differences in the method adopted by each author for measuring equivalent width, so some mismatches are to be expected. In particular \citet{lucky13} measured the equivalent width of the DIB 5780 as a sum of a narrow and a wide component, so we do refrain from comparison of her DIB 5780 with our. For the same reason we excluded DIB 6202 from the measurements for HD 30614. A large difference for DIB 5705 in HD 30614 between \citet{thorburn03} and other sources is probably a measuring error, because our measurement and the one from JD94 agree very well. In the end we retain 5 measurements in 4 different stars. Table \ref{tab:match} gives values measured in this paper and those from the literature. Figure \ref{fig:match} shows a graphical representation of accepted comparisons from table \ref{tab:match}. The best linear fit is
\begin{equation}
\begin{array}{ll}W_{this\ paper}=&(1.035 \pm 0.036)\cdot W_{literature}\\&+(1.85 \pm 1.72) m\mathrm{\AA,}\end{array}
\end{equation}
and the correlation coefficient is 0.96. The relation between equivalent widths from this paper and from the literature agrees well with the null hypothesis.

\subsection{Reddening}

Reddening in this paper is always represented by color excess E(B-V). Most of the reddening data (for exact numbers see table \ref{tab:list}) is taken from \citet{savage85}, where reddening was derived from Johnson UBV and Cape U$_c$BV photometry. Uncertainties of reddening for individual sightlines are not given, but are of the order of 0.03 magnitude ($\pm 2 \sigma$). The second most used source was \cite{snow77}. This is a combined catalogue of many different surveys of DIBs and includes reddening for most of the sightlines. Several different sources were used and the paper gives no uncertainties to reddening data. For the first observing session we used reddening from \citet{maeder75} where several B-type stars with well known reddening are listed. Reddening for two very well studied stars were taken from \citet{slyk06}. Some of the data was taken from \citet{valencic04}. This catalogue, made from UV and IR photometry, includes reddening and total extinction in V band as well as uncertainties, that are around 0.04 mag for E(B-V) for most of the stars. Last reddening source was \citet{neckel80}. Extinction was calculated from UBV and H$\beta$ photometry. Only total extinction in the V band is given, but it was calculated from measured reddening with a total to selective extinction coefficient of 3.1 for all sightlines, so the reddening can be recalculated without loss of information. Typical errors in E(B-V) are of the order of 0.05 magnitude.

Reddening in some sightlines has more than one reference in literature. In these cases the average value was used.
\\
\section{Correlations and groups of DIBs}

\label{sec:common}

\subsection{Correlations}

Calculation of the correlation coefficient:
\begin{equation}
\rho_{x,y}=\frac{\mathrm{cov}(x,y)}{\sigma _x \sigma _y}
\end{equation}
is the most basic measure of dependence of the two quantities $x$ and $y$. Equivalent widths of DIBs are expected to correlate with each other and with reddening and abundances of some other atoms, ions and molecules. We measured parameters of interstellar lines of Na {\sc i} (5891.6 \AA\ and 5897.6\AA), Ca {\sc i} (4227.9 \AA), Ca {\sc ii} (3934.8 \AA\ and 3969.6 \AA), CH (4300.3 \AA), CH$^+$ (4232.3 \AA) and CN (3874.6 \AA). All the correlation coefficients of measured equivalent widths and reddening are in table \ref{tab:corr}.

Correlations with sodium and calcium lines must be treated separately, because they are saturated even in sightlines with low reddening. Relation from \citet{munari97} for Na D line was used to modify measured equivalent widths into a linear scale. Correlations were then calculated with calculated column densities. For Ca {\sc i} and Ca {\sc ii} lines, relations from \citet{welty03} were used. Molecular lines are weak enough to assume a linear relation between column density and equivalent width.

Several pairs of DIBs show a very high correlation value of 0.9 or more. This could suggest that these DIBs have the same carrier, but correlation does not necessary imply causation.

Let us take Na lines and Ca {\sc ii} line at 3969 \AA\ for example. Two Na lines correlate very well with each other (correlation coefficient is 0.97) but they also correlate well with Ca {\sc ii} line (correlation coefficients are 0.89 and 0.81). All the correlations with reddening are worse than that. A conclusion that Na and Ca {\sc ii} lines have the same carrier is obviously false. Therefore a better way to determine common carriers is needed.

\subsection{Common carriers identification}
\label{sec:c}

If we have a correlation coefficients for correlation between one DIB and reddening and another DIB and reddening, statistical limits for correlation coefficient between both DIBs can be calculated \citep{olkin81}.

We take a correlation matrix for two DIBs ($A$ and $B$) and reddening $R$:
\begin{equation}
\mathcal{C}= \left[
\begin{array}{ccc}
1 & \rho & \sigma\\
\rho & 1 & \tau\\
\sigma & \tau & 1\\
\end{array}\right],\quad
\begin{array}{l}
\rho=\mathrm{corr}(A,B),\\
\sigma=\mathrm{corr}(A,R),\\
\tau=\mathrm{corr}(B,R),
\end{array}
\end{equation}
where Greek letters are correlation coefficients between every pair of quantities. A correlation matrix has a property that its determinant must always be larger or equal to zero. This property can be transformed into inequality
\begin{equation}
\sigma \tau-\sqrt{(1-\sigma^2)(1-\tau^2)}\leq \rho \leq \sigma\tau + \sqrt{(1-\sigma^2)(1-\tau^2)},
\end{equation}
which are essentially lower and upper limits for $\rho$. This equation is symmetric for any of the three correlation coefficients.

If two DIBs correlate well with each other, but each of them correlates well with reddening, it cannot be concluded that they have the same carrier. They could as well have two different carriers that correlate with reddening in the same manner.

When we calculate both limits, it is possible to estimate which pairs of DIBs are more probable to have the same carrier. This will be the ones that correlate much better with each other than with reddening. We assume an expected correlation ($\rho_e$) to be in the middle of the lower ($\rho_l$) and upper ($\rho_u$) limit for $\rho$. We introduce a factor $\mathcal{P}$, which should increase with the probability that two DIBs have the same carrier. $\mathcal{P}$ will be called probability factor and is equal to: 
\begin{equation}
\mathcal{P}=(\rho_c-\rho_e)\rho_c,
\label{eq:p}
\end{equation}
where $\rho_c$ is the calculated correlation factor. The difference between calculated and expected correlation factor is multiplied by calculated correlation factor, because we are only interested in pairs with high correlation. High correlation and large positive differences imply a higher probability that the two DIBs share the same carrier.

Let us return to the example with Na and Ca {\sc ii} lines. Probability factor that two Na lines have the same carrier is 0.36 and for two Ca {\sc ii} lines it is 0.38. Probability factors that Na and Ca {\sc ii} lines have the same carrier are between 0.09 and 0.15.

This method will not identify ions and will not distinguish between two different elements with very similar influence on reddening. However this method performs well on observed interstellar elements and is worthy to be tried on DIBs. Because the carriers of the DIBs are not known, we will compare the results with correlation plots in section \ref{sec:r2}.

\section{DIBs and reddening}

\label{sec:r1}

\begin{figure*}[!ht]
\centering
\includegraphics[width=\textwidth]{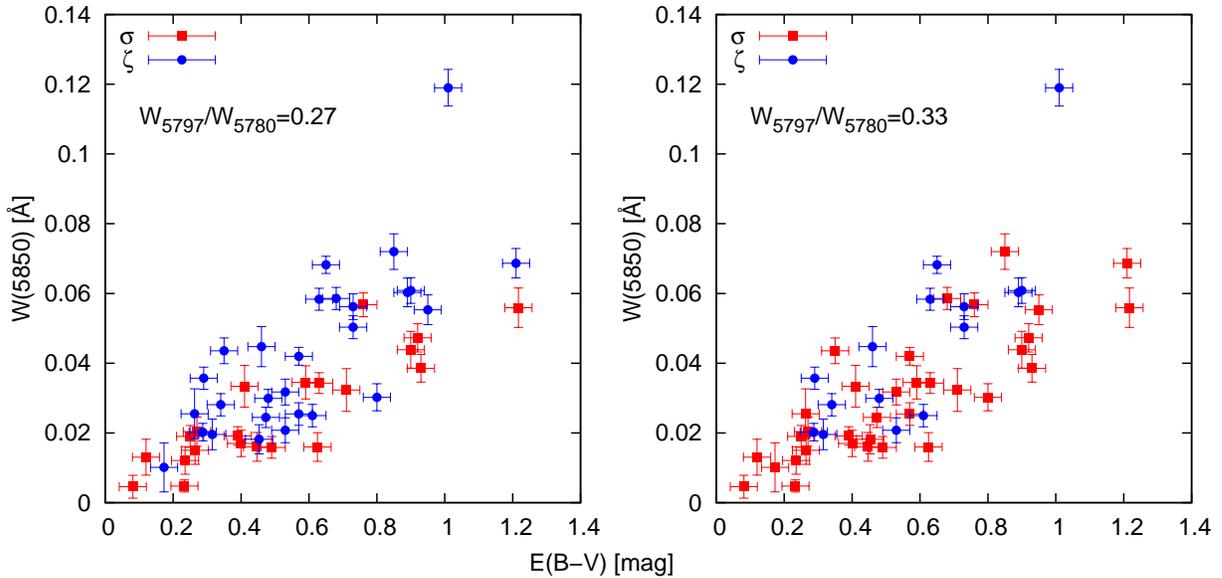}
\caption{Changing the boundary condition between $\sigma$ (black,red) and $\zeta$ (gray, blue) sightlines to 0.27 (left) and 0.33 (right). The two sightlines do never form two completely separate groups. Color version of this figure is available only in the electronic edition.}
\label{fig:boundary}
\end{figure*}

Table \ref{tab:corr} shows that most of DIBs correlate well with reddening. However the correlations are not perfect. An obvious scatter is present, with reduced $\chi^2$ distinctively larger than 1. Reduced $\chi^2$ is computed as described in \citet{nrc} with method \textsc{chixy} and linear fits (where used) were calculated as described in method \textsc{fitexy}. 

We put a lot of care into the error estimation. Scatter of measurements in the correlation plots is not due to errors but shows varying conditions of the ISM.

\section{$\sigma$ and  $\zeta$ sightlines}

\label{sec:r2}

\begin{figure*}[!p]
\centering
\includegraphics[width=\columnwidth, height=5.7cm]{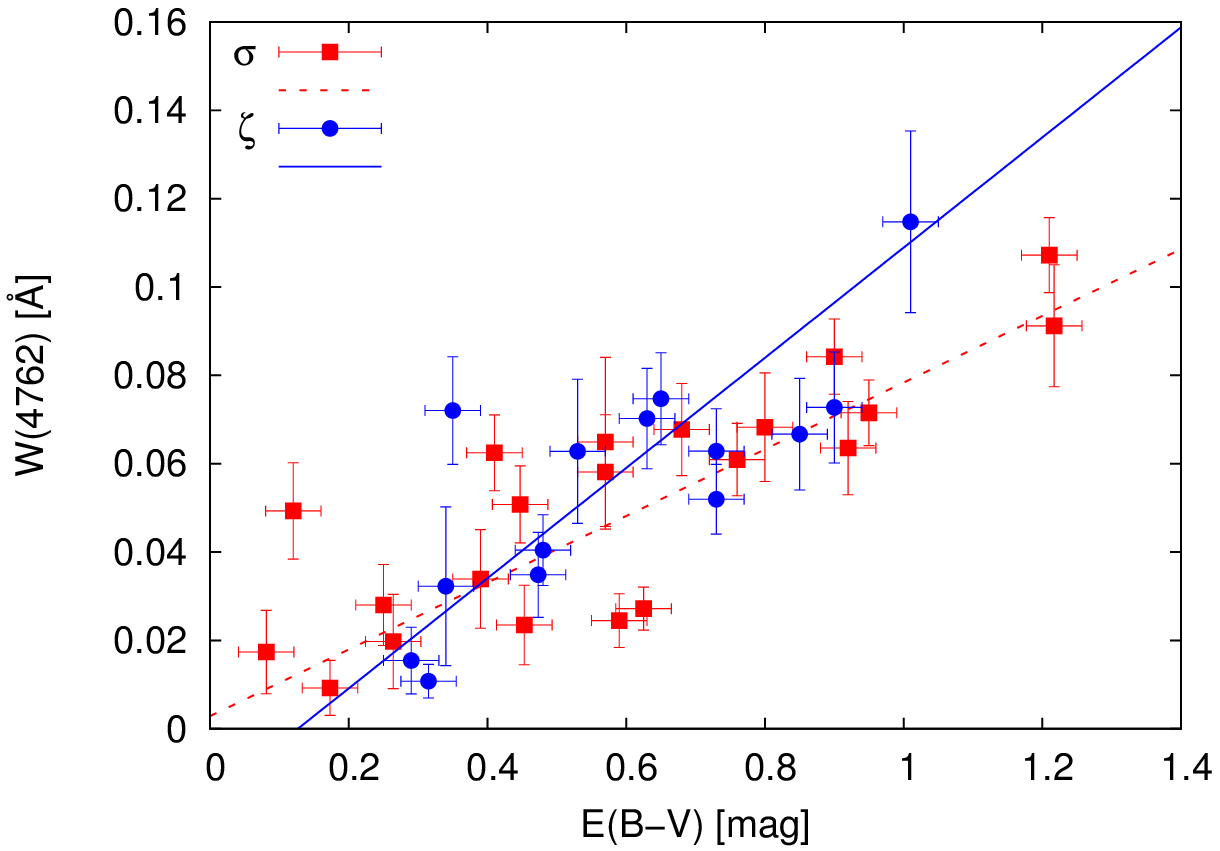}
\includegraphics[width=\columnwidth, height=5.7cm]{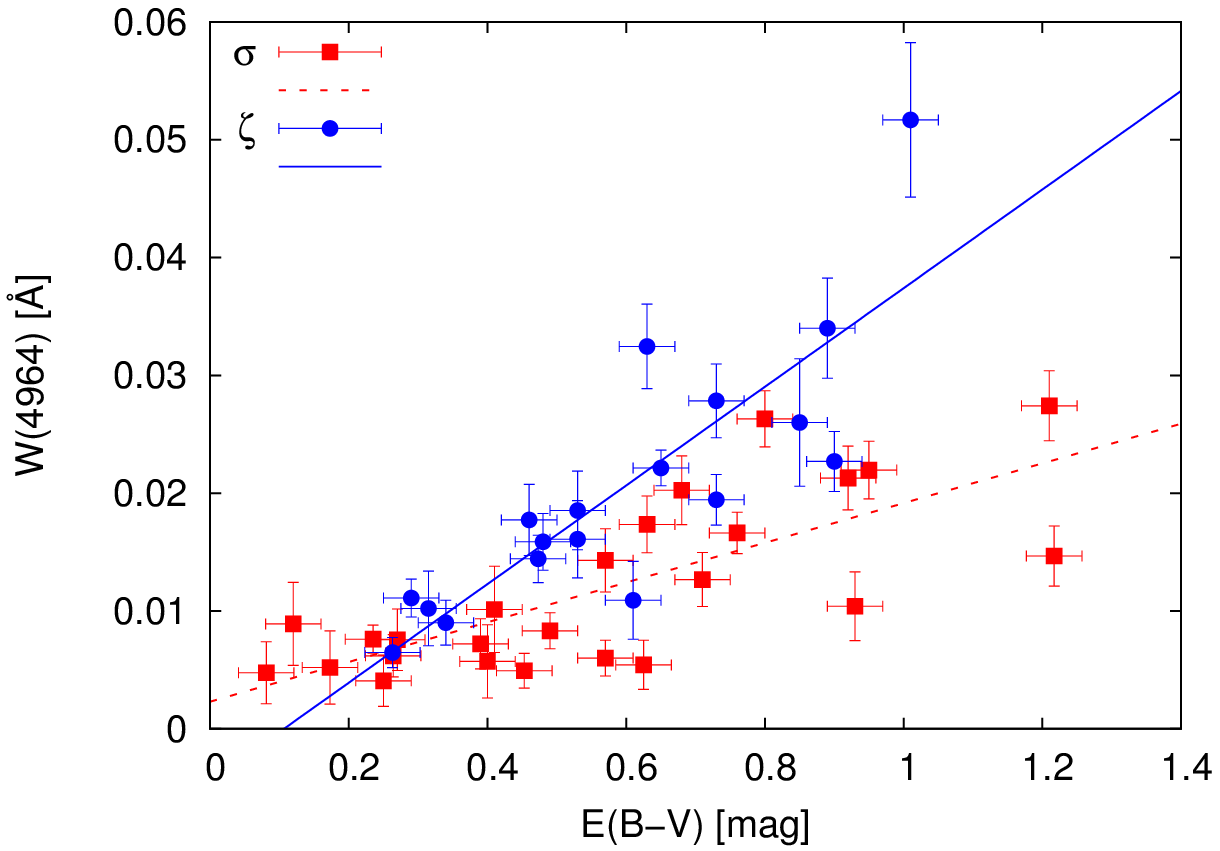}\\
\includegraphics[width=\columnwidth, height=5.7cm]{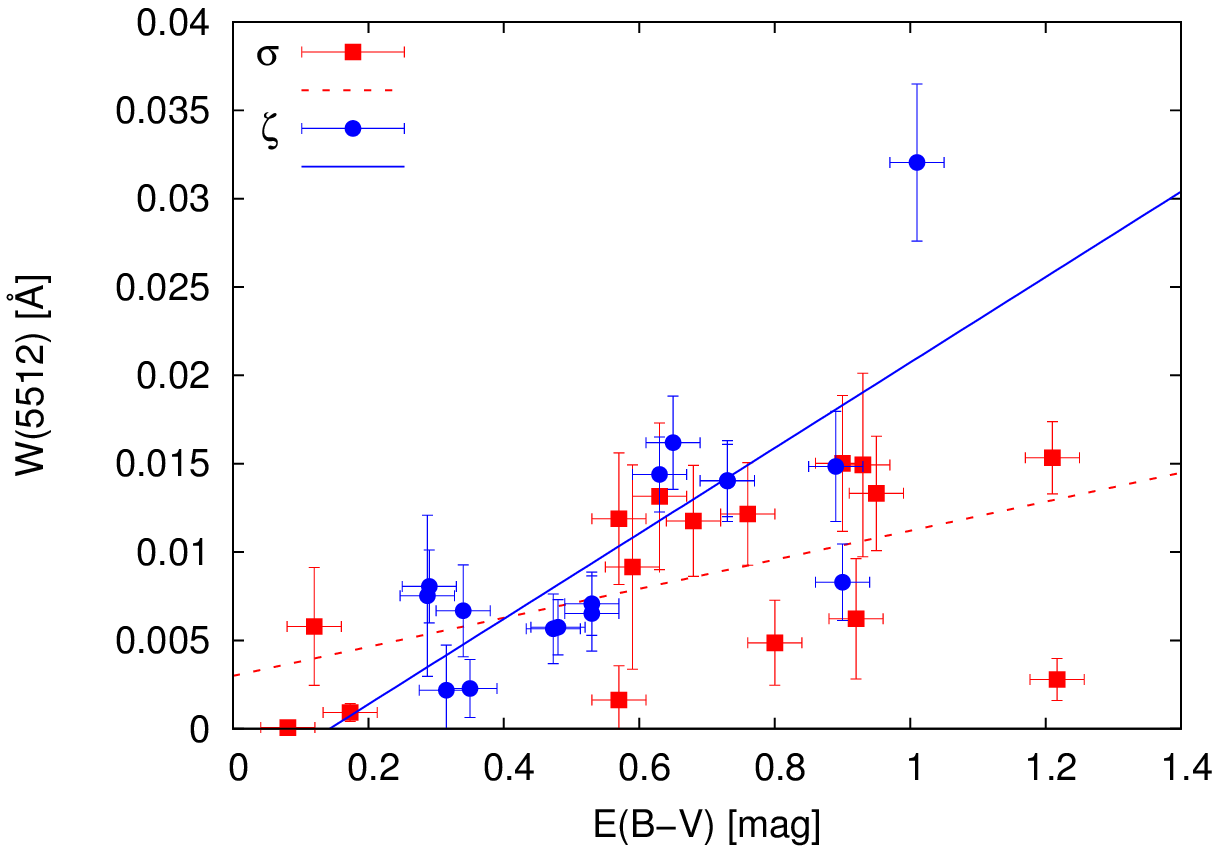}
\includegraphics[width=\columnwidth, height=5.7cm]{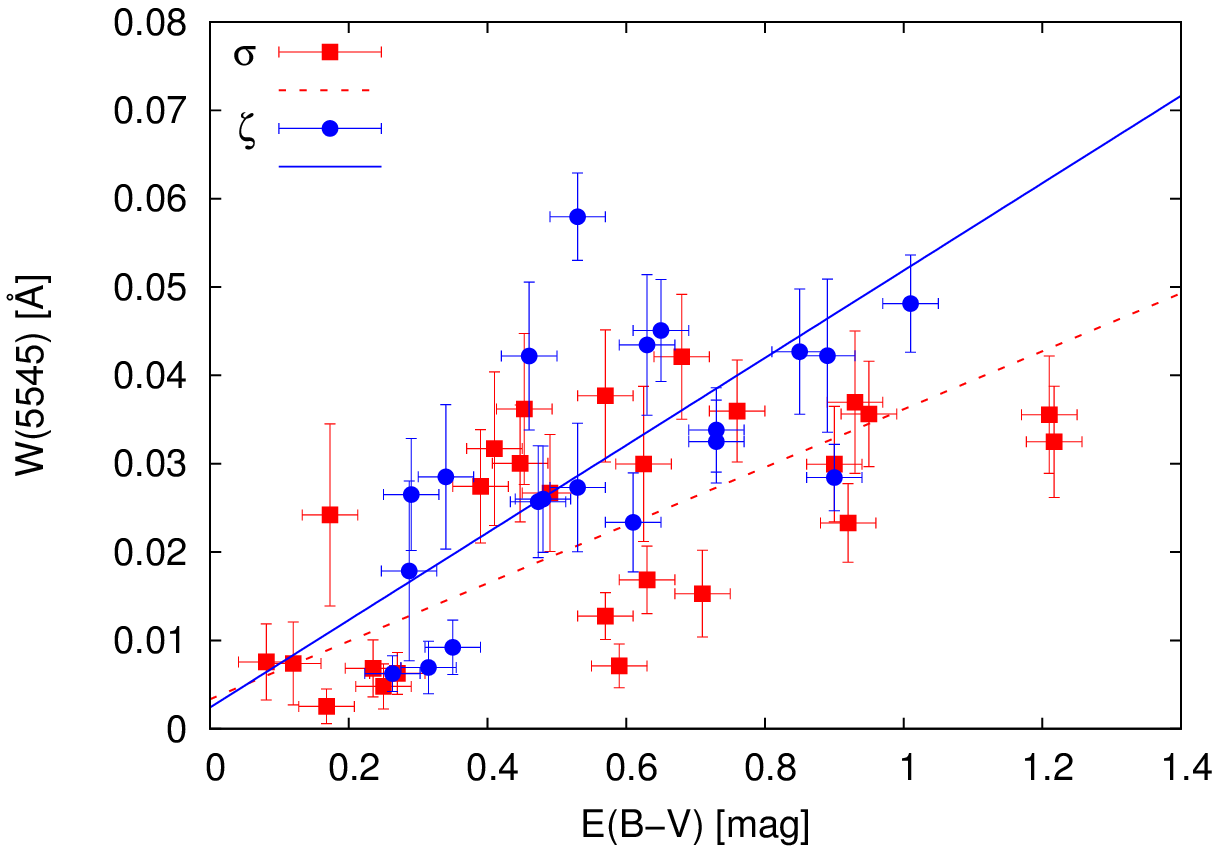}\\
\includegraphics[width=\columnwidth, height=5.7cm]{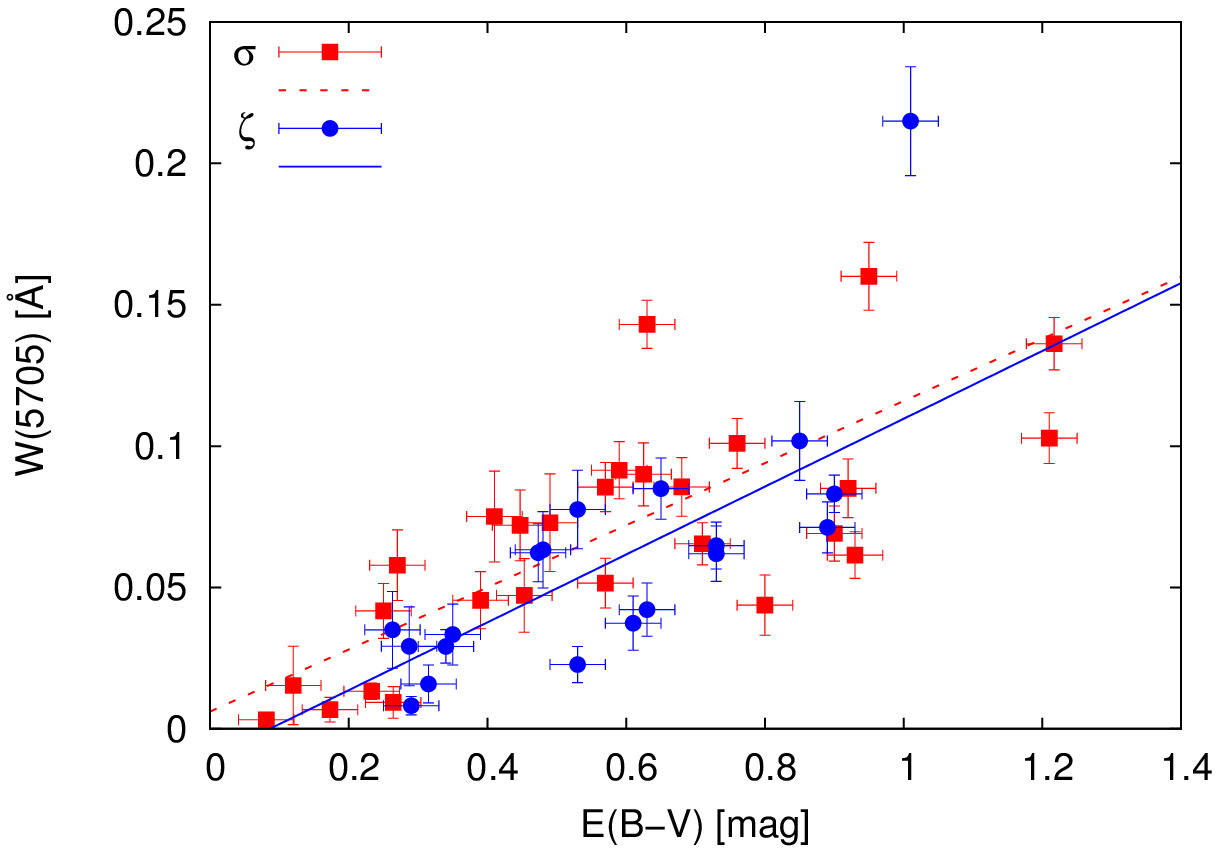}
\includegraphics[width=\columnwidth, height=5.7cm]{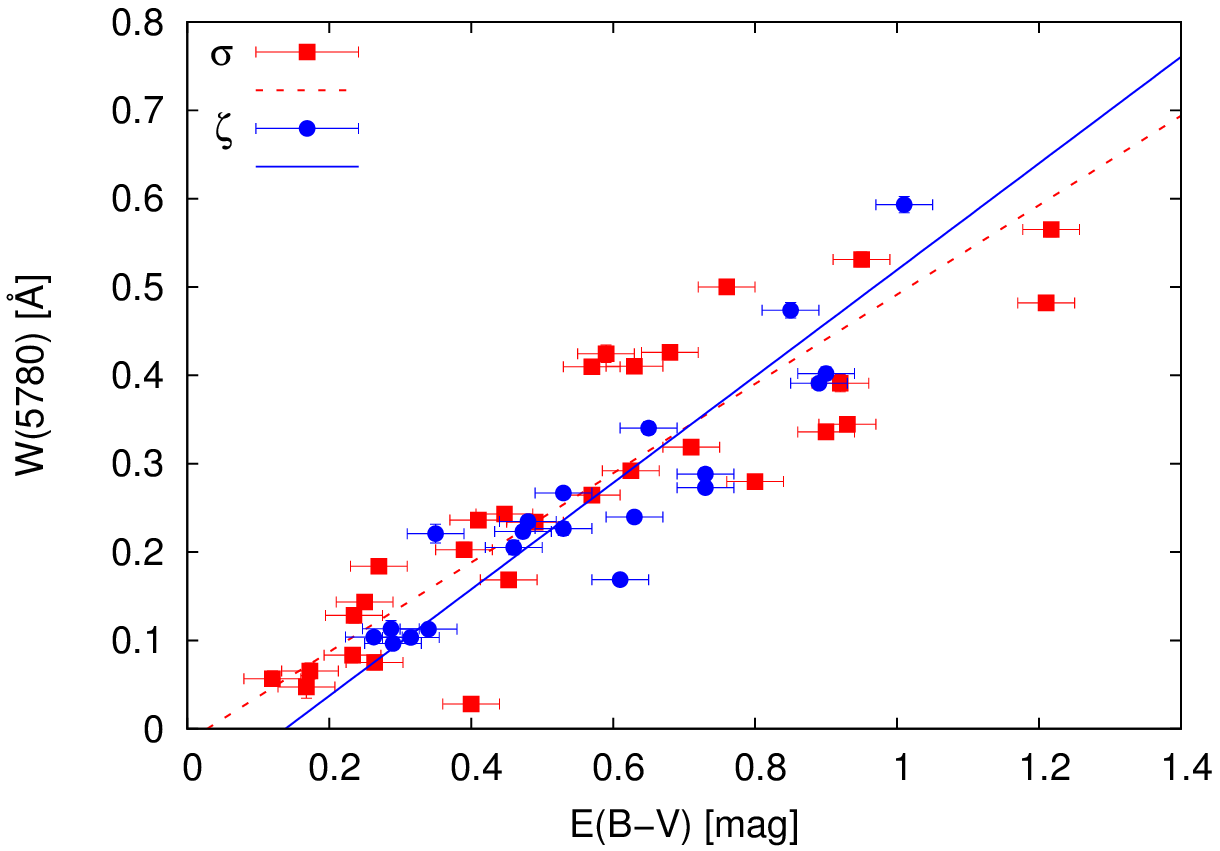}\\
\includegraphics[width=\columnwidth, height=5.7cm]{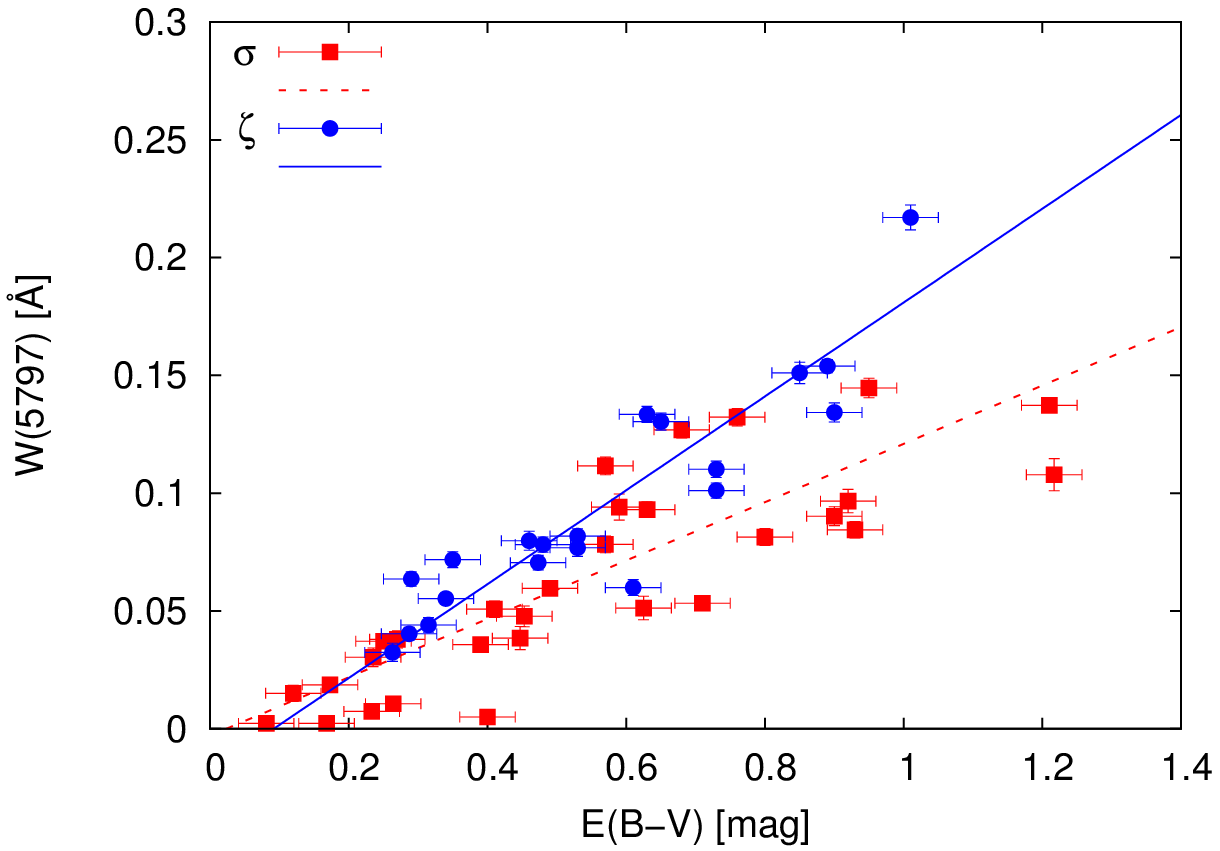}
\includegraphics[width=\columnwidth, height=5.7cm]{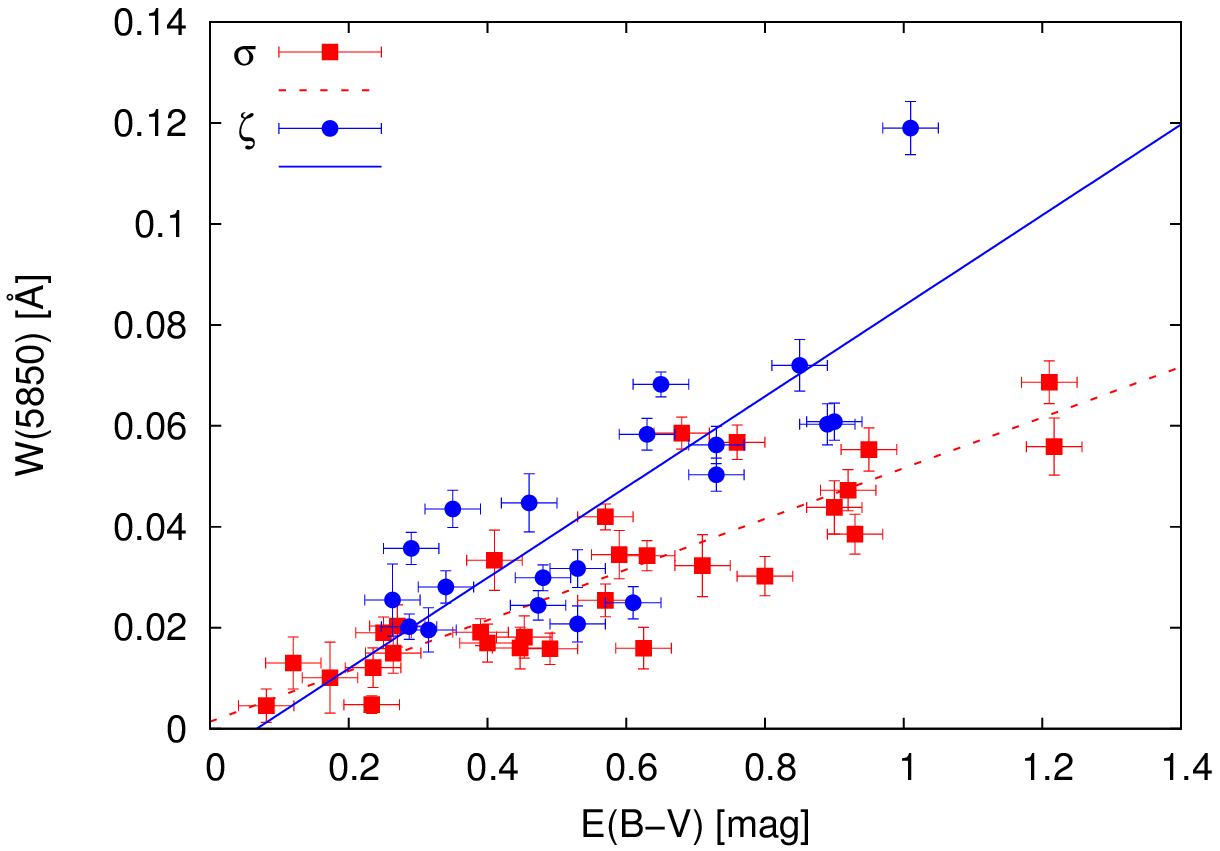}\\
\label{fig:corelations}
\caption{Correlations between equivalent width of DIBs and reddening. Lines were fitted to relations for $\sigma$ (black or red dashed line, black or red square points) and $\zeta$ (gray or blue solid line, gray or blue circular points) sightlines. Parameters of linear fits are given in table \ref{tab:rel}. Color version of this figure is available only in the electronic edition.}
\label{fig:cor}
\end{figure*}
\begin{figure*}[!h]
\setcounter{figure}{10}
\centering
\includegraphics[width=\columnwidth, height=5.7cm]{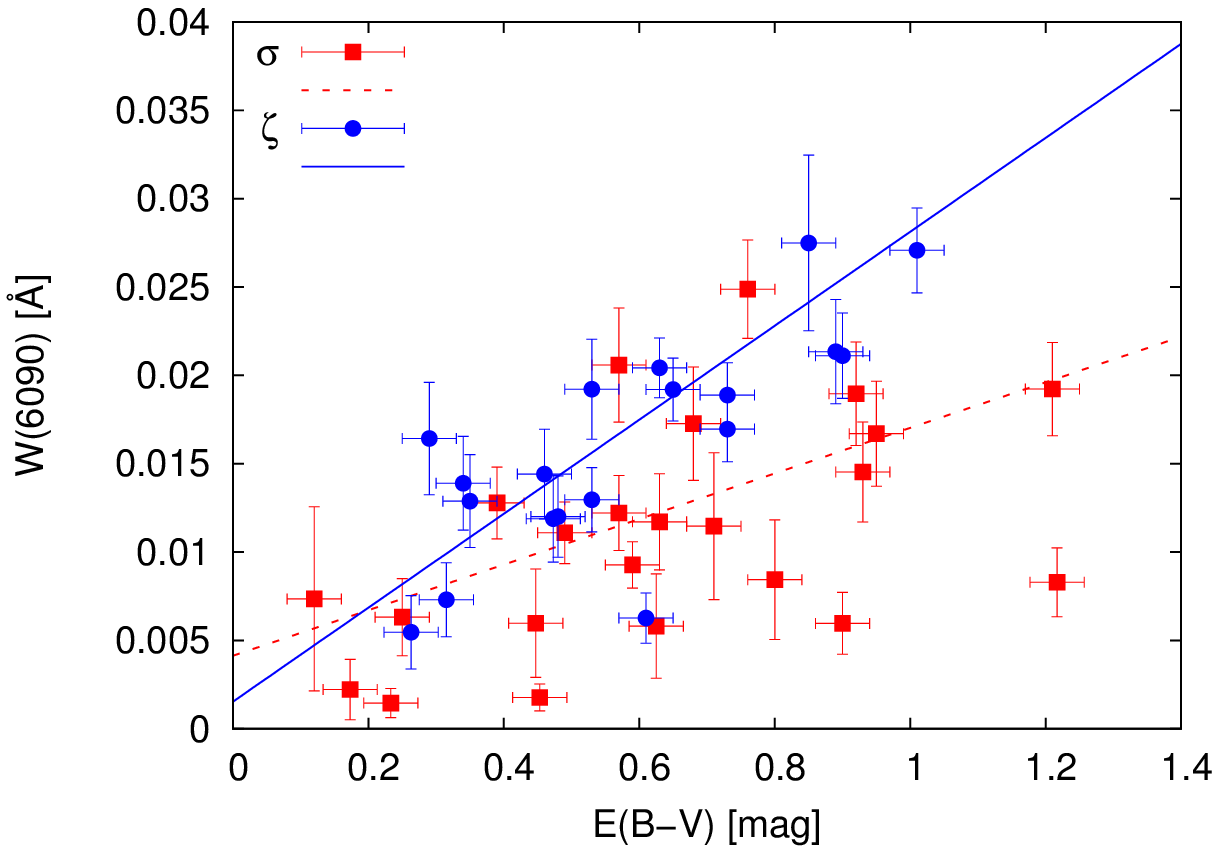}
\includegraphics[width=\columnwidth, height=5.7cm]{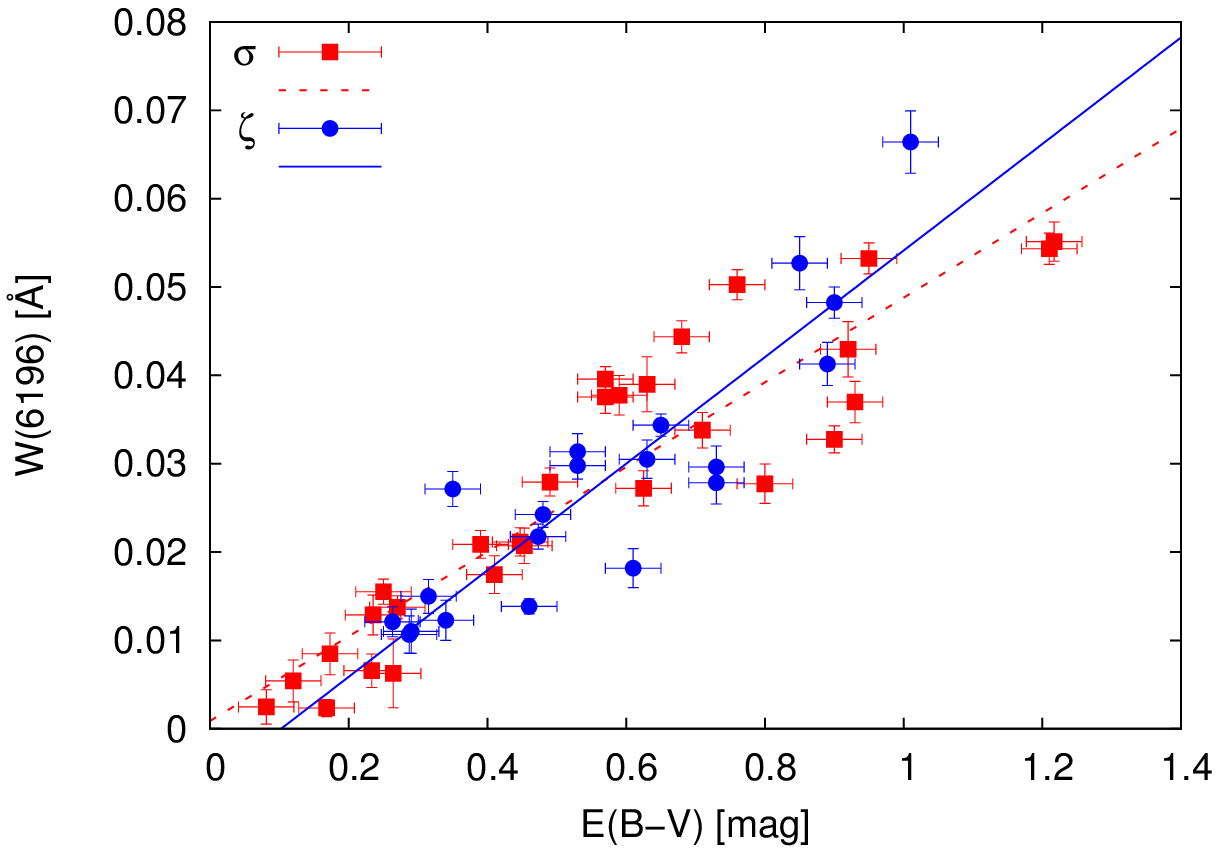}\\
\includegraphics[width=\columnwidth, height=5.7cm]{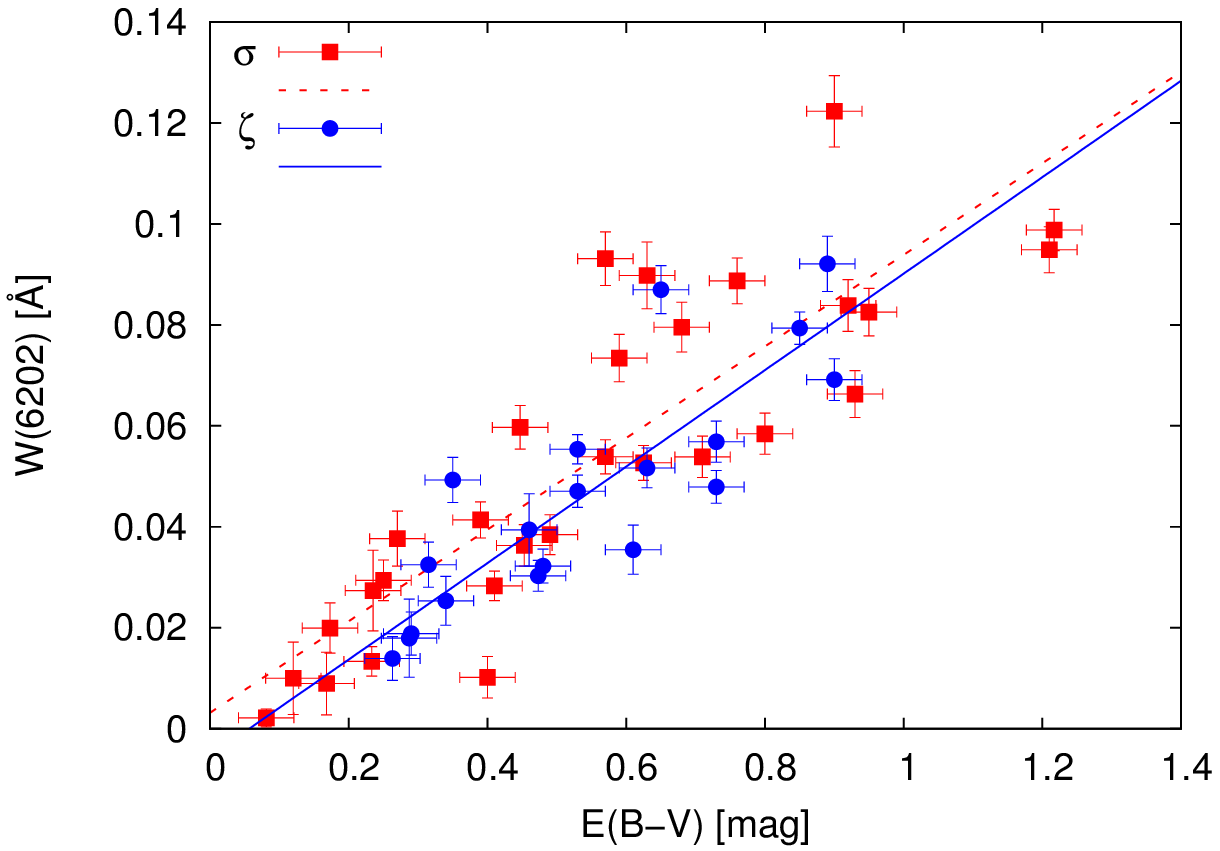}
\includegraphics[width=\columnwidth, height=5.7cm]{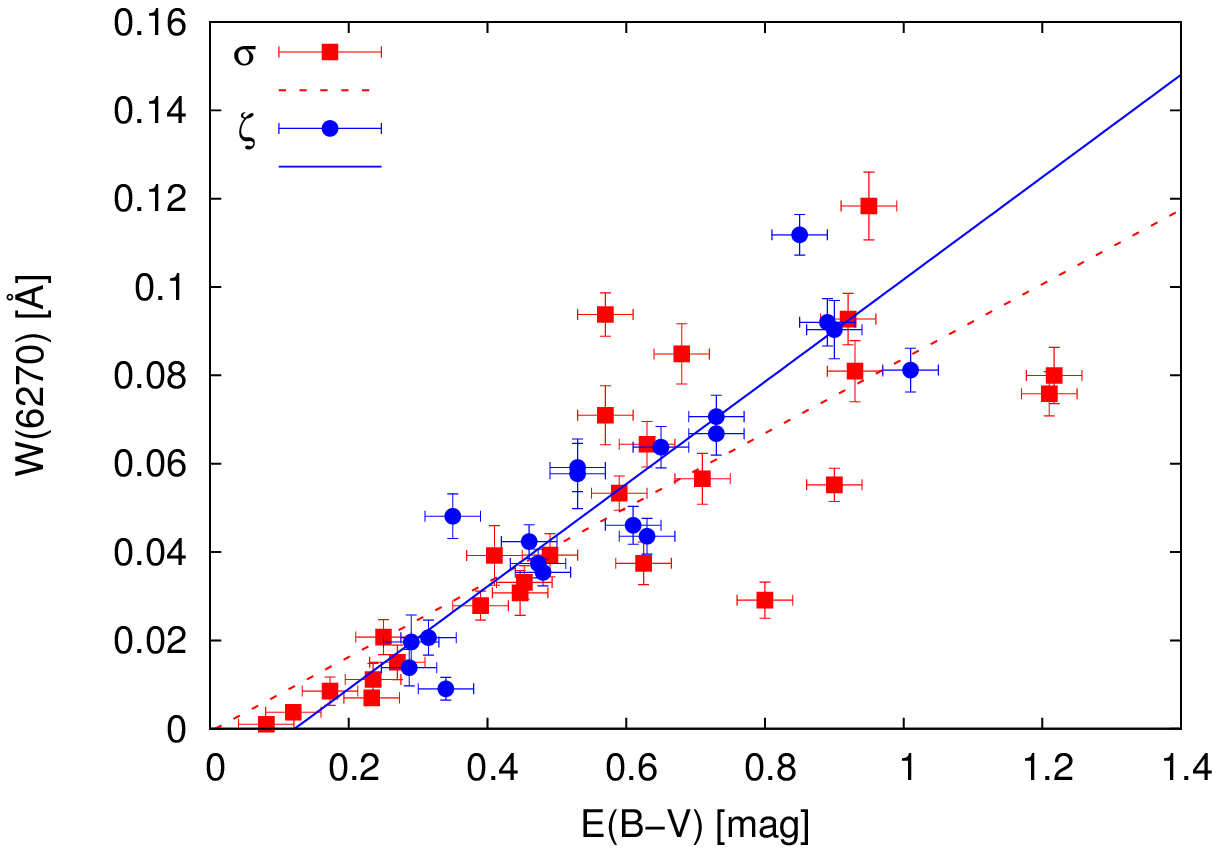}\\
\includegraphics[width=\columnwidth, height=5.7cm]{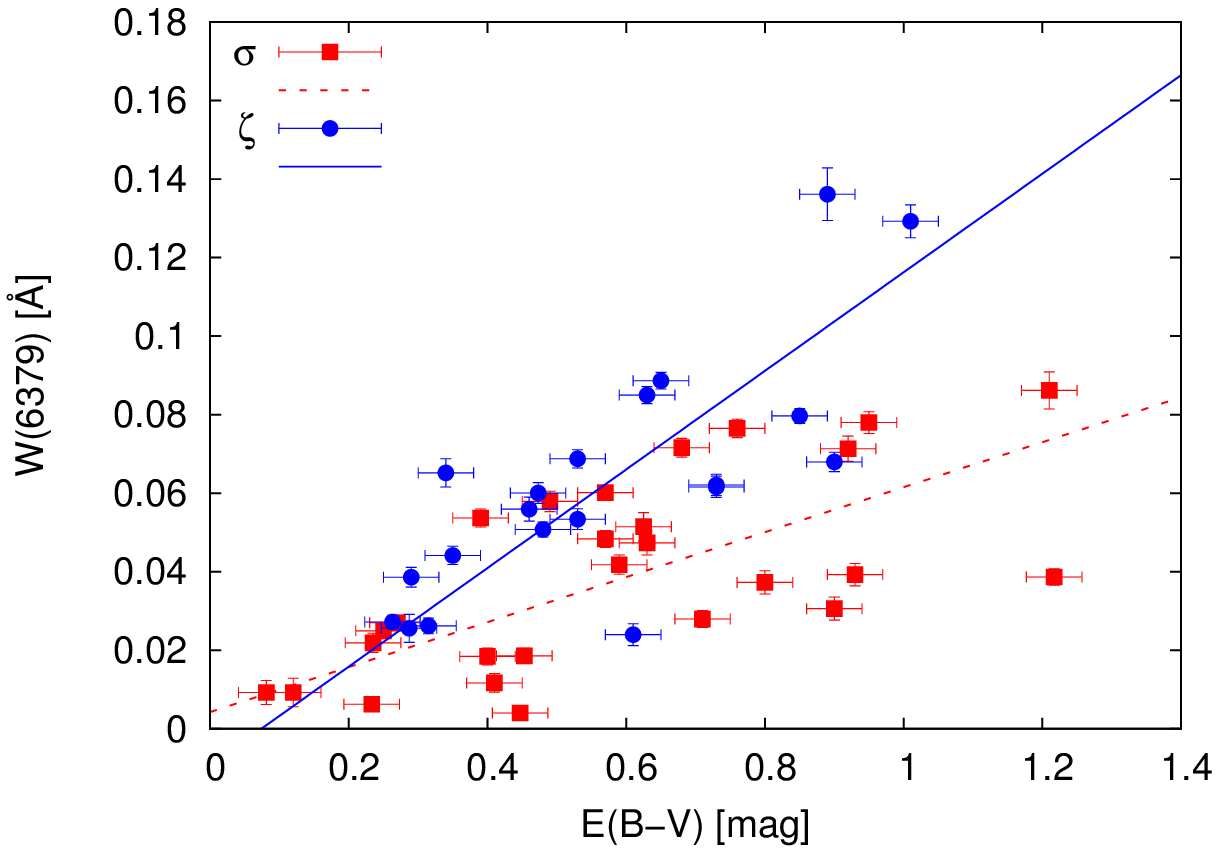}
\includegraphics[width=\columnwidth, height=5.7cm]{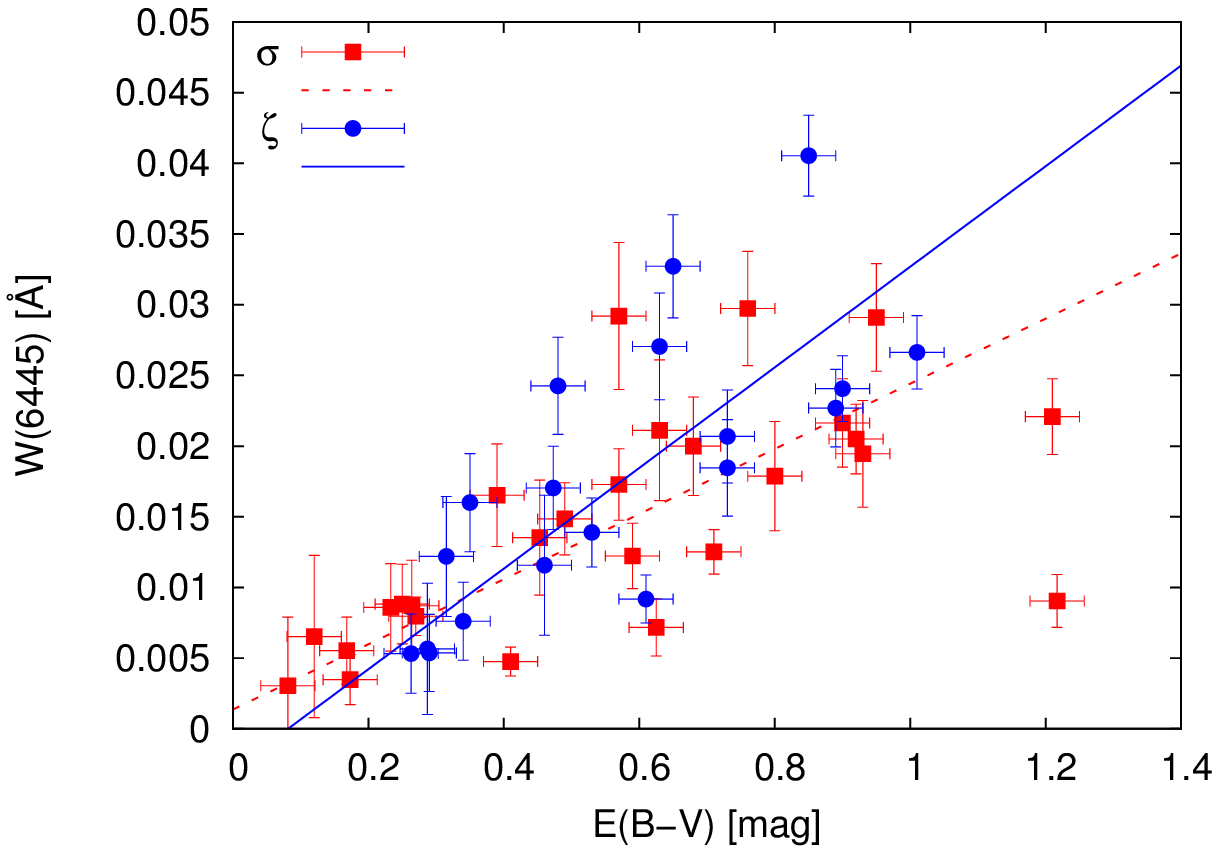}\\
\includegraphics[width=\columnwidth, height=5.7cm]{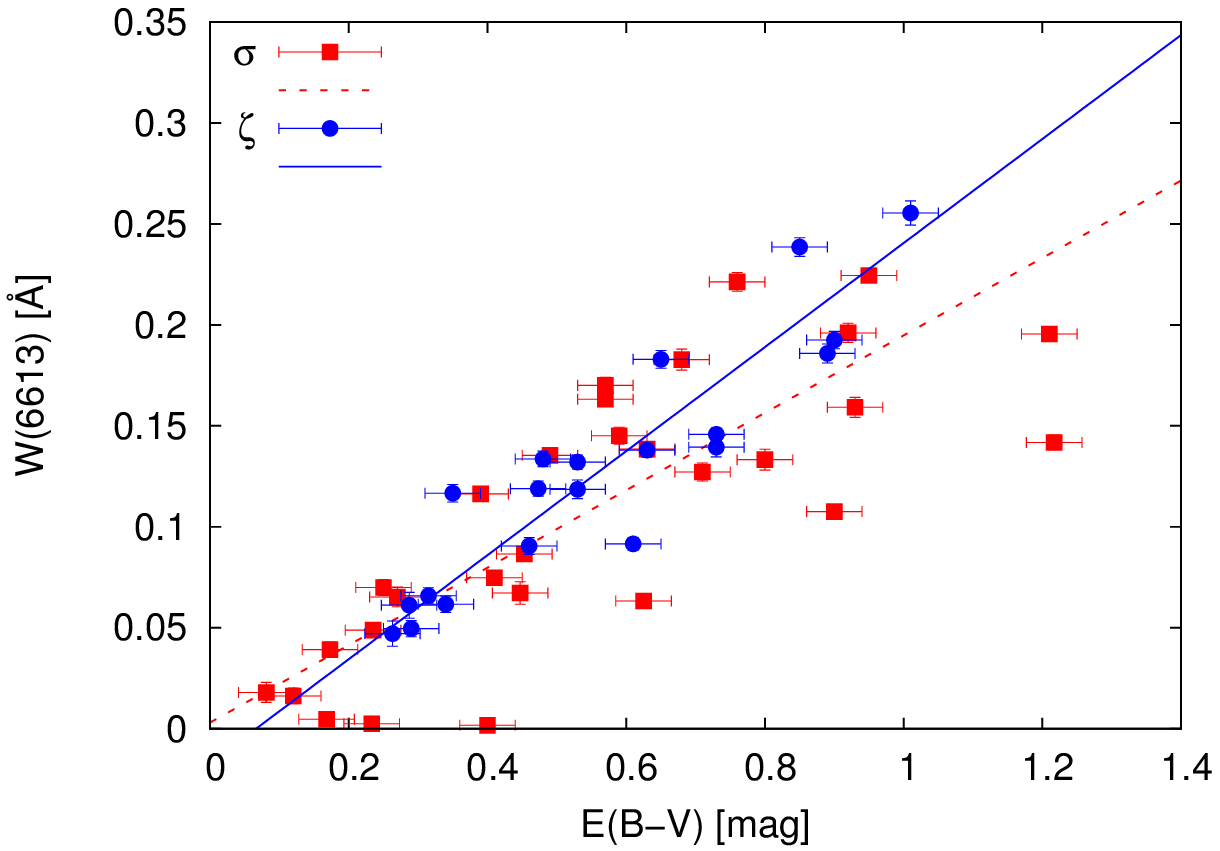}
\includegraphics[width=\columnwidth, height=5.7cm]{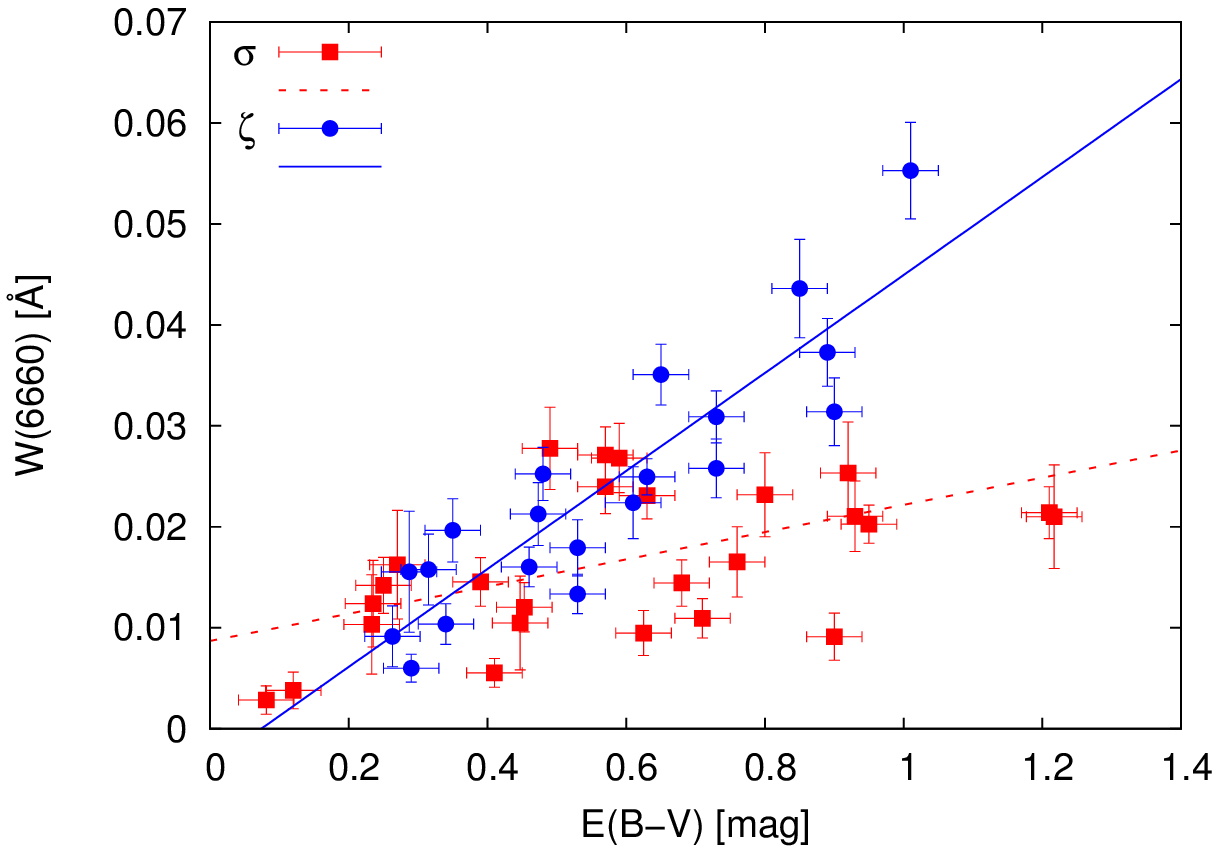}\\
\caption{Continued.\\[0.5cm]}
\end{figure*}

\citet{krelowski92} showed that parts of ISM that are shielded from the UV radiation field produce different ratios between equivalent widths of DIB 5780 and 6797. UV shielded sightlines are named type $\zeta$ after a sightline toward star $\zeta$ Oph and non-shielded sightlines are named type $\sigma$ after $\sigma$ Sco. Intermediate sightlines must also exist so the transition between two types is smooth. Original definition is based on the central depth of the two DIBs, where the boundary ratio of the central depths $A_{5797}/A_{5780}$ equals 0.4. This corresponds to approximately 0.3 for the ratio of equivalent widths. Due to a smooth transition, the exact boundary ratio is not very important. We show in figure \ref{fig:boundary} that points denoted as $\sigma$ and $\zeta$ sightlines form a mixed group without a sharp boundary, even when the adopted value for the boundary ratio is varied by 10\%\ in either direction. There are always some points representing $\zeta$ sightlines in the area dominated by $\sigma$ sightline points and vice versa. 

\begin{figure}[!ht]
\centering
\includegraphics[width=\columnwidth]{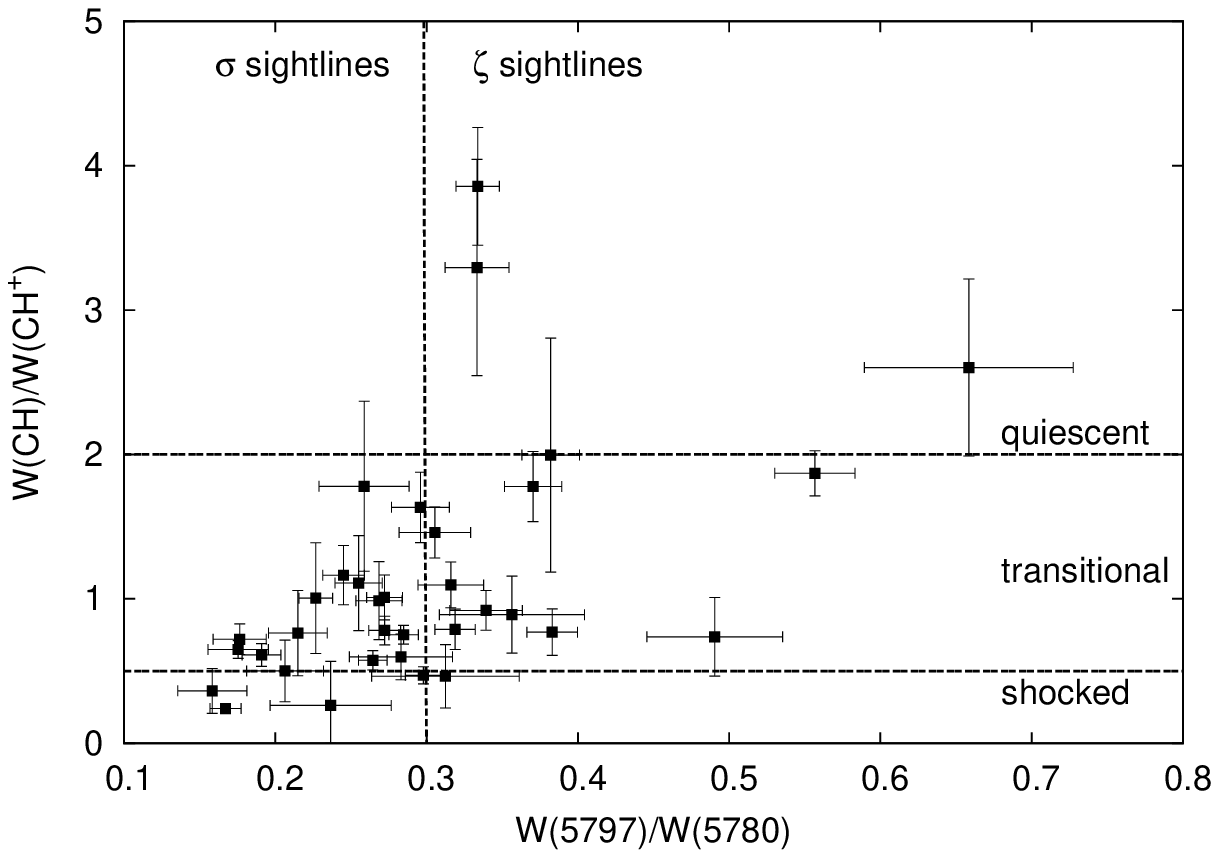}
\caption{Stars plotted in an diagram showing different states of the ISM. Vertical line divides sightlines into $\sigma$ and $\zeta$ types. Horizontal lines divide them into shocked and turbulent part and quiescent part. There is a wide transitional area where most of the sightlines lie.}
\label{fig:diag}
\end{figure}

Existence of sightlines with different DIB ratios also reflects in the equivalent width -- reddening relation as shown in \citet{vos11} for DIB 5780. The relation from \citet{vos11} is based on the observations of a small area in Scorpius and gives a different relation than this study, that has the distribution of sightlines along a much larger part of the galactic plane. Figure \ref{fig:cor} shows correlations between reddening and equivalent width of 16 DIBs separatelly for $\sigma$ and $\zeta$ sightlines. 

From figure \ref{fig:cor} it is evident, that the linear relation for $\zeta$ sightlines is significantly steeper in some cases than the relation for $\sigma$ sightlines. Higher UV shielding in the centres of the interstellar clouds can explain this, if the rate of the dissociation of the DIB carrier molecules due to the UV radiation is lower in the centres than in the outer unshielded parts. This results in a higher density of DIB carriers per unit of reddening. 

Linear fit for the $\zeta$ sightlines does not reach zero equivalent width at zero reddening, but at $E(B-V)\simeq0.09$ magnitude. This is due to the fact, that UV shielded regions can only exist where dust density is high enough. This value seems to be characteristic to all observed DIBs.

Based on the slopes of the linear fits, DIBs from figure \ref{fig:cor} can be divided into two types. 5705, 5780, 6196, 6202 and 6270 do not show much difference between $\zeta$ and $\sigma$ sightlines. We will call this group of DIBs type {\sc i}. Slopes of the two linear relations are close to each other and correlation in $\sigma$ sightlines is slightly lower than correlation in $\zeta$ sightlines. DIBs 4964, 5797, 5850, 6090, 6379 and 6660 have distinctively different relations for $\zeta$ and $\sigma$ sightlines. We will call this group type {\sc ii} DIBs. The correlation between reddening and equivalent widths for this type of DIBs is significantly better in $\zeta$ sightlines than in $\sigma$ sightlines.

In general, the correlation between reddening and equivalent width is better in $\zeta$ sightlines with few statistically insignificant exceptions. The correlation coefficients are listed in the table \ref{tab:rel}. $\chi^2$ test further confirms this with reduced $\chi^2$ $\sim 30$ \% smaller for $\zeta$ sightlines than for $\sigma$ ones. However the reduced $\chi^2$ still does not reach values close to 1.

This distinction between two types of DIBs is also predicted by a method from section \ref{sec:c}. Two types of DIBs can be recognized in table \ref{tab:prob}. DIBs that belong to the same type all share higher probability factors than DIBs of the other type or the uncategorised DIBs. For example, the probability factor between every pair of DIBs 4964, 5797, 5850, 6379 and 6660 are all between 0.21 and 0.30. This is considered to be a high probability factor and implies a common carrier. All these are type {\sc ii} DIBs. Probability factors between these 5 DIBs and type {\sc i} DIBs, like 5705, 5780 and 6202 are in the range from 0.03 to 0.12, which is significantly lower.

We also checked for similar behaviour due to electron density, gas density and turbulence or shock wave presence. 

Turbulence or slow shockwaves can be detected trough CH$^+$ and CH ratio \citep{gredel02}. One of the most important mechanisms of CH$^+$ creation involves shockwaves with minimal velocity of 8 km/s. High N(CH$^+$)/N(CH) thus implies presence of shockwaves. There are 3 sightlines that show shocks and 3 that appear to be quiescent. Others lie in between, where neither shock or quiescent state can be confirmed (figure \ref{fig:diag}). A mild correlation between ratios of W(5797)/W(5780) and  N(CH$^+$)/N(CH) is observed in our sample of ISM spectra, implying that shockwaves are more likely present in the ISM exposed to the UV light.

\begin{figure}[!ht]
\centering
\includegraphics[trim=0cm 0cm 0cm 0cm, width=\columnwidth]{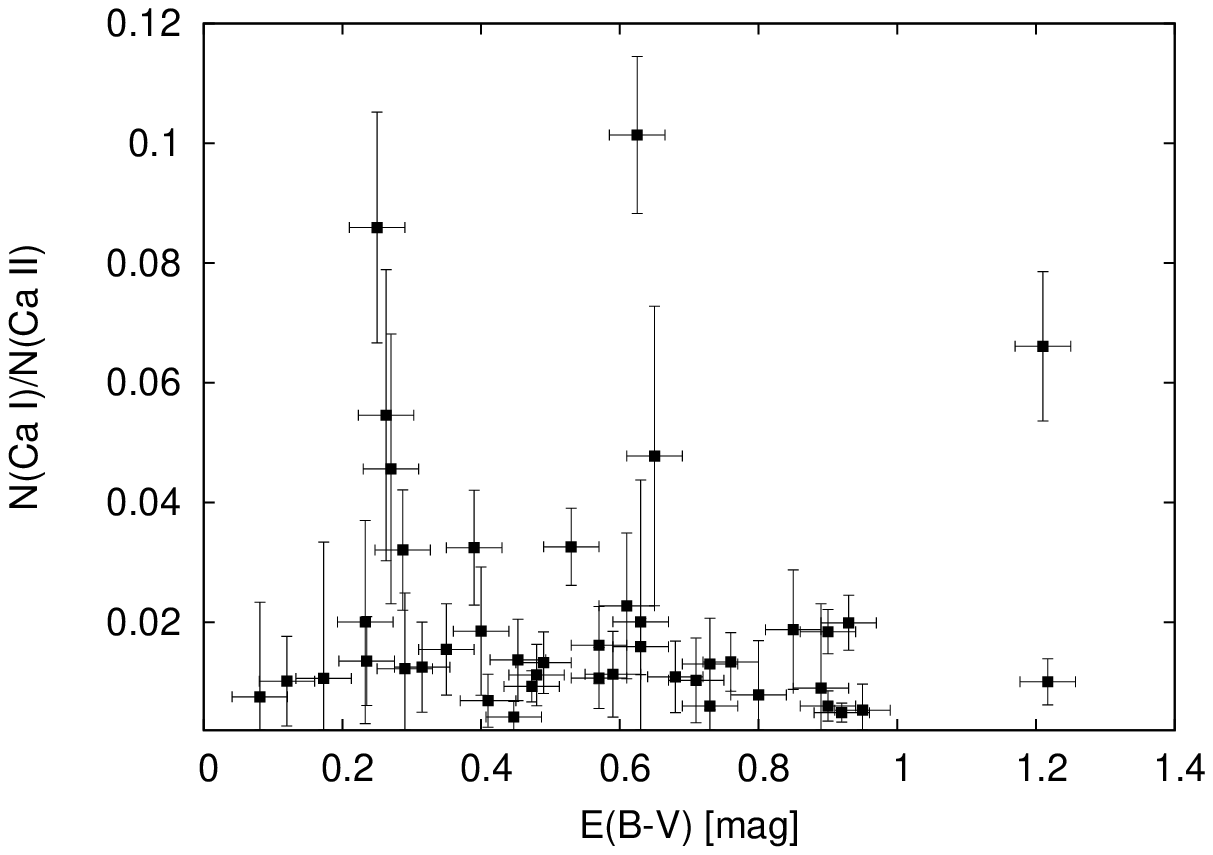} \includegraphics[trim=-0.8cm 0cm 0cm 0cm, width=\columnwidth]{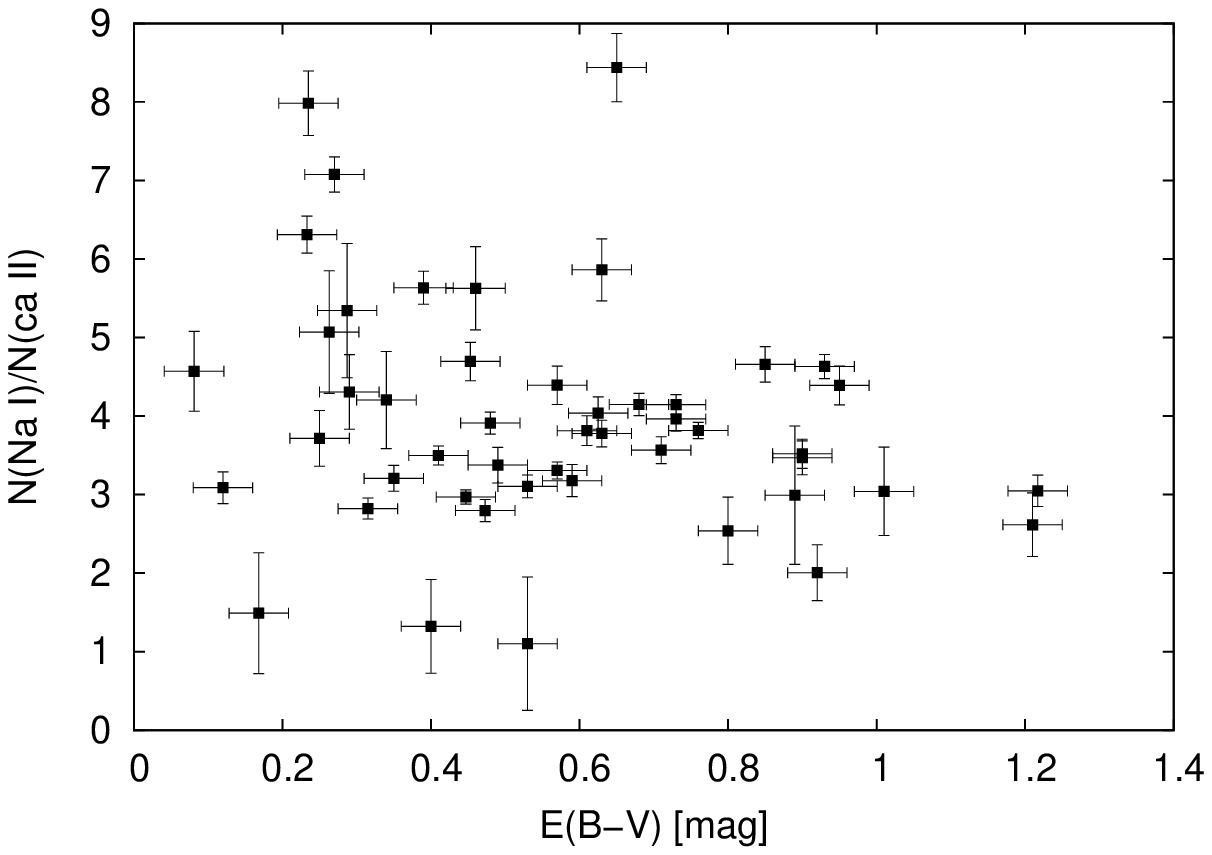}
\caption{N(Ca {\sc i})/N(Ca {\sc ii}) is proportional to electron density if there is no Ca {\sc ii} deposition on-to dust grains going on. This can be tested trough Na {\sc i} to Ca {\sc ii} ratio, which is very sensitive to Ca {\sc ii} deposition. values between 2 and 10 are common. Values close to 100 would indicate Ca {\sc ii} deposition on grains and values below 1 would indicate a presence of strong shockwaves that break dust grains and release Ca {\sc ii}. There was no deposition or release of Ca {\sc ii} observed. Electron densities are therefore in the range from $\sim$0.2 to $\sim$15 $e^-/cm^{3}$}
\label{fig:ostalo}
\end{figure}

Electron density can be calculated trough the ratio of Ca {\sc i} or Ca {\sc iii} and Ca {\sc ii} lines. Whenever we detect Ca {\sc i} lines, it can be assumed that Ca {\sc iii} lines are absent. This is important, because Ca {\sc iii} has no strong lines in the observed spectral range. Electron density is calculated as \citep{draine}:
\begin{equation}
n_e=\frac{\sigma(\mathrm{Ca\ {\textsc{i}}})}{\sigma(\mathrm{Ca\ {\textsc{ii}}})}\left( \frac{\Gamma}{\alpha}\right)_{\mathrm{Ca\ {\textsc{i}}}},
\end{equation}
where $\sigma$ is a column density.  $\left( \frac{\Gamma}{\alpha}\right)$ is a ratio of a photoionization rate and a radiative recombination coefficient and is in the order of $60\ cm^{-3}$\citep{peq86} for ISM expected in the observed sightlines and does not vary significantly when changing temperature or density. Electron density measured from N(Ca\ {\textsc{i}})/N(Ca\ {\textsc{ii}}) is therefore reliable if no deposition of Ca {\sc ii} on the grains is present \citep{welty03}. Presence of deposition can be evaluated trough N(Na\ {\textsc{i}})/N(Ca\ {\textsc{ii}}) \citep{ritchey06}. Figure \ref{fig:ostalo} shows N(Ca {\sc i})/N(Ca {\sc ii}) and N(Na {\sc i})/N(Ca {\sc ii}) ratios.

\begin{figure}[!ht]
\centering
\includegraphics[width=\columnwidth]{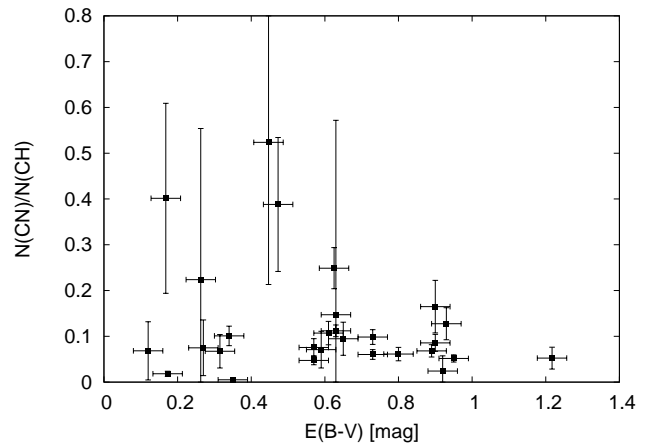}
\caption{$N(CN)/N(CH)$ traces the gas density. Observed ratios correspond to densities between 95 and 750 cm$^{-3}$, approximately.}
\label{fig:density}
\end{figure}

Gas density (figure \ref{fig:density}) is proportional to N(CN)/N(CH) \citep{pan05}. Because the chemistry of molecules is much more diverse, a precise relation cannot be established \citep{craw90}. However a correlation exists and approximate relation to transform ratios into density is \citep{pan05}:
\begin{equation}
n=\frac{N(CN)}{N(CH)}*960\ cm^{-3}.
\end{equation}

No behaviour similar to UV shielding was observed when separating sightlines based on turbulence (figure \ref{fig:diag}), electron density (figure \ref{fig:ostalo}) or gas density (figure \ref{fig:density}). 

\section{Conclusions and discussion}
Above results show, that modelled asymmetric Gaussians fit very well to DIB profiles in spectra with a resolution power of 23 000 and a high signal to noise ratio. However, asymmetric Gaussians are purely artificially constructed functions and do not resemble any physical properties of DIBs. It is well known that DIB profiles have fine substructure, but at lower resolutions the profiles can be fit with simple functions. In addition, asymmetric Gaussians seem to fit all observed DIBs adequately. Estimated errors are lower than errors given by a simple integration of observed pixel fluxes. Fitting also yields line widths and asymmetries that are otherwise hard to measure at a comparable precision.

Correlation of measured equivalent widths was observed for all pairs of spectral lines and reddening. As it is dangerous to make conclusions about common carriers from DIBs alone, we always compared correlations of three quantities, accounting for a statistical probability how two correlation coefficients influence the third one. This is important, because correlation is imperfect not only due to errors of the measured equivalent width but also due to different behaviour of DIBs in different sightlines. DIBs identified to be similar with this method, also appear to have a similar relation when separating sightlines into $\sigma$ and $\zeta$ types in the reddening versus DIB strength plots.

Separating $\sigma$ and $\zeta$ sightlines based on the ratio of equivalent widths of DIB 5797 to DIB 5780 proved to be successful. Sightlines of both types are coherently distributed on the sky, where the sightlines toward the same OB association belong to the same type.  Similar, the nearby sightlines most often belong to the same type as the light penetrates the same complex of interstellar clouds. Observed sightlines are divided almost in half, what is consistent with the original condition, based on line depths. The results are insensitive  to the adopted borderline value dividing the two groups. Results are consistent when varying the boundary condition for $\pm$~10\% around the chosen value of 0.3.

Separating $\sigma$ and $\zeta$ sightlines shows at least two different types of DIBs. We introduce type {\sc i} DIBs where the behaviour of DIB equivalent width and reddening is similar for $\sigma$ and $\zeta$ sightlines, and type {\sc ii} DIBs where it differs significantly. This separation clearly improves correlation coefficients (compared to the correlation before the separation into $\sigma$ and $\zeta$ sightlines) and $\chi^2$ for $\zeta$ sightlines in a type {\sc ii} DIBs, where DIB equivalent width vs. reddening relation is significantly different for both types of sightlines. There is no or only small improvement of the correlation coefficient and $\chi^2$ for $\zeta$ sightlines in type {\sc ii} DIBs. The correlation for $\sigma$ sightlines in type {\sc i} and type {\sc ii} DIBs is not, or only insignificantly improved. This means that UV shielding is one of the dominant factors of the DIB equivalent width vs. reddening relation shape for $\zeta$ sightlines, but in $\sigma$ sightlines other physical parameters play a major role. Complementary explanation could be a skin effect that affects type {\sc ii} DIBs and weakens the DIBs in $\sigma$ sightlines. See also a discussion in \citet{adamkovics05}.

Typical type {\sc i} DIBs are 5705, 5780, 6196, 6202 and 6270. Typical type {\sc ii} DIBs are 4964, 5797, 5850, 6090, 6379, 6613 and 6660. The remaining 4 DIBs are hard to associate with any of these two types.

Different behaviour of $\sigma$ and $\zeta$ sightlines in type {\sc ii} DIBs reflects in a different steepnes of slopes of the relation between the DIB's equivalent width and reddening. The change of slope is most obvious when comparing DIBs 6613 and 6660. For DIB 6613 the two slopes have a ratio of $1.24\pm 0.22$ and for DIB 6660 it is $3.59\pm1.19$, which is almost 2 sigmas apart. This could indicate that there are more than two types of DIBs. Additional observations will further test this hypothesis.

All results are listed in tables in section \ref{sec:tables}. Table \ref{tab:list} lists basic properties of observed sightlines and the quality of the recorded data. Table \ref{tab:corr} gives correlation coefficients for all pairs of the observed interstellar lines and reddening. In a similar format, the probability factors (defined in equation \ref{eq:p}) are given in table \ref{tab:prob}. Table \ref{tab:rel} provides linear relations, correlation coefficients and reduced $\chi^2$ for reddening vs.\ DIB's equivalent width. Table \ref{tab:example} serves as an example of what parameters are published in the electronically provided table for each of 50 sightlines.

Na {\sc i} and Ca {\sc ii} were detected in all sightlines, Ca {\sc i} and CH in 46, CH$^+$ in 45 and CN in 31 sightlines, including lower density $\sigma$ sightlines. Ca {\sc i} and molecular lines are weak and all lie in blue part of the spectral range, where SNR drops to around 200 or less. Therefore the conditions of the ISM cannot be measured precisely, especially where one of the lines is expected to be very weak. There is also not much variation in the conditions of the ISM. No difference in DIB -- reddening relation, when separating sightlines with respect to ISM conditions, is thus expectable. More sightlines would be needed to sample more diverse ISM clouds in order to be able to separate DIB -- reddening relations based on other conditions, in addition to UV shielding. With our sample of observed sightlines, we were unable to separate sightlines based on turbulence, electron density or gas density.
\paragraph{Acknowledgements}
We thank the anonymous referee and the editor for valuable comments and suggestions which improved the quality of the paper and its presentation.
We thank Ulisse Munari and the time allocation committee of the Asiago observatory for allocating extensive amount of observing time and observatory staff for a valuable assistance during observations. TZ acknowledges support of the Slovenian Research Agency and Centers of excellence programmes.\\

\bibliographystyle{apj}
\bibliography{dibbib}

\section{Tables}
\label{sec:tables}

\begin{table*}[!h]
\centering
\caption{Basic properties of observed stars. Notes: Observing session codes: 1 (1. 16. 2011), 2 (9. 15. 2011 -- 9. 16. 2011), 3 (2. 3. 2012 -- 2. 6. 2012), 4 (5. 3. 2012 -- 5. 5.2012), 5 (8. 29. 2012 -- 9. 5. 2012), 6 (24. 11. 2012 -- 26. 11. 2012). References: v: \citet{valencic04}, s: \citet{snow77}, a: \citet{savage85}, m: \citet{maeder75}, l: \citet{slyk06}, n: \citet{neckel80}. Coordinates, magnitudes and spectral types are taken from Simbad.}
\begin{tabular}{c c c c c c c c c}
\hline\hline
Name & l & b & Spectral Type & V & E(B-V) & Distance$^\mathrm{n}$ & Observing Session & Peak {S$/$N} \\\hline
 &$^\circ$ & $^\circ$& & mag & mag & kPc & & \\\hline
HD 167451 & 16.8262 &  1.5203 & B1I & 8.30 & 1.01$^\mathrm{n}$ & 1.540 & 2 & 480\\
HD 176853 & 24.5945 &  -7.3127 & B2V & 6.64 & 0.46$^\mathrm{n}$ & 0.305 & 2 & 470\\
HD 192660 & 77.3737 &  3.1375 & B0I & 7.56 & 0.90$^\mathrm{v,a}$ & 1.447 & 2 & 510\\
HD 212571 & 66.0068 &  -44.7395 & B1V & 4.79 & 0.233$^\mathrm{s}$ & 0.319 & 2 & 530\\
HD 206165 & 102.2713 &  7.2469 & B2I & 4.79 & 0.473$^\mathrm{a}$ & 0.576 & 2 & 510\\
HD 190918 & 72.6516 &  2.0651 & W-R & 6.81 & 0.447$^\mathrm{s}$ & 1.015 & 2 & 540\\
HD 193183 & 75.9500 &  1.4992 & B1.5I & 7.02 & 0.625$^\mathrm{s}$ & 2.491 & 2 & 530\\
HD 194839 & 79.5171 &  1.8727 & B0.5I & 7.53 & 1.217$^\mathrm{a}$ & 1.000 & 2 & 490\\
HD 203025 & 97.9974 &  6.5304 & B2III & 6.43 & 0.453$^\mathrm{s}$ & 0.481 & 2 & 500\\
HD 205139 & 100.5455 &  6.6217 & B1II & 5.54 & 0.39$^\mathrm{s}$ & 0.763 & 2 & 540\\
HD 209744 & 103.2795 &  3.4975 & B1V & 6.71 & 0.48$^\mathrm{s}$ & 0.554 & 2 & 510\\
HD 222568 & 116.5030 &  6.3694 & B1IV & 7.71 & 0.65$^\mathrm{n}$ & 0.772 & 2 & 510\\
HD 5005 & 123.1232 & -6.2436 & O5.5 & 7.76 & 0.41$^\mathrm{s}$ & 2.262 & 2 & 520\\
HD 10516 & 131.3247 &  -11.3301 & B2V & 4.09 & 0.264$^\mathrm{m}$ & 0.320 & 1 & 540\\
HD 22951 & 158.9198 &  -16.7030 & B0.5V & 4.98 & 0.263$^\mathrm{m}$ & 0.402 & 1 & 420\\
HD 24131 & 160.2266 &  -15.1369 & B1V & 5.78 & 0.27$^\mathrm{m}$ & 0.554 & 1 & 390\\
HD 25539 & 163.5201 &  -14.6703 & B3V & 6.87 & 0.287$^\mathrm{m}$ & 0.415 & 1 & 420\\
HD 25833 & 163.2454 &  -13.7164 & B5V & 6.69 & 0.235$^\mathrm{m}$ & ? & 1 & 430\\
HD 45995 & 200.6231 &  0.6784 & B2V & 6.14 & 0.168$^\mathrm{m}$ & 0.410 & 1 & 380\\
HD 161056 & 18.6702 & 11.5808 & B1.5V & 6.30 & 0.63$^\mathrm{a}$ & 0.167 & 3 & 550\\
HD 22253 & 144.2776 &  0.9244 & B0.5III & 6.545 & 0.61$^\mathrm{a}$ & 0.718 & 3 & 560\\
HD 23180 & 160.3637 &  -17.7398 & B1III & 3.85 & 0.29$^\mathrm{v,l}$ & 0.289 & 1, 3 & 520\\
HD 24398 & 162.2892 &  -16.6904 & B1I & 2.883 & 0.34$^\mathrm{l}$ & 0.352 & 1, 3 & 560\\
HD 25638 & 143.6705 &  7.6576 & B0III & 6.93 & 0.73$^\mathrm{v,s}$ & 1.085 & 3 & 610\\
HD 25639 & 143.6755 &  7.6583 & B0III & 6.86 & 0.73$^\mathrm{s, a}$ & 1.085 & 3 & 520\\
HD 35921 & 172.7615 &  0.6106 & O9.5III & 6.85 & 0.63$^\mathrm{n}$ & 1.562 & 3 & 490\\
HD 137569 & 21.8661 & 51.9303 & B5III & 7.91 & 0.40$^\mathrm{a}$ & 0.564 & 4 & 460\\
HD 176304 & 42.8531 &  2.8820 & B2V & 6.75 & 0.49$^\mathrm{a}$ & 0.317 & 4 & 490\\
HD 207260 & 102.3099 &  5.9342 & A2I & 4.31 & 0.53$^\mathrm{n}$ & 1.069 & 4 & 500\\
HD 207308 & 103.1089 &  6.8176 & B0.5V & 7.49 & 0.53$^\mathrm{s}$ & 0.765 & 4 & 500\\
HD 217086 & 110.2206 &  2.7197 & O7V & 7.71 & 0.95$^\mathrm{v,a}$ & 0.980 & 4 & 510\\
HD 217312 & 110.5631 &  2.9460 & B0IV & 7.44 & 0.68$^\mathrm{a}$ & 0.972 & 4 & 520\\
HD 193443 & 76.1497 &  1.2833 & O9III & 7.24 & 0.71$^\mathrm{a}$ & 1.238 & 5 & 520\\
HD 194279 & 78.6780 &  1.9863 & B1.5I & 7.09 & 1.21$^\mathrm{a}$ & 0.970 & 5 & 510\\
HD 210839 & 103.8281 &  2.6107 & O6I & 5.08 & 0.57$^\mathrm{a}$ & 0.550 & 5 &530\\
HD 216898 & 109.9274 &  2.3930 & O8.5V & 8.04 & 0.85$^\mathrm{a}$ & 1.394 & 5 & 500\\
HD 217035 & 110.2532 &  2.8609 & B0V & 7.74 & 0.76$^\mathrm{a}$ & 0.880 & 5 & 510\\
HD 217297 & 110.8177 &  3.5221 & B1.5V & 7.41 & 0.57$^\mathrm{s,a}$ & 0.774 & 5 & 540\\
HD 217919 & 111.2657 &  3.3094 & B0IV & 8.24 & 0.93$^\mathrm{a}$ & 0.894 & 5 & 500\\
HD 218323 & 111.7993 &  3.7279 & B0III & 7.65 & 0.90$^\mathrm{a}$ & 1.083 & 5 & 510\\
HD 218376 & 109.9477 & -0.7834 & B0.5IV & 4.84 & 0.25$^\mathrm{v,a}$ & 0.382 & 5 & 520\\
HD 209975 & 104.8707 &  5.3906 & O9.5I & 5.11 & 0.35$^\mathrm{a, v}$ & 0.992 & 6 & 550\\
HD 14956 & 135.4234 & -2.8622 & B2I & 7.24 & 0.89$^\mathrm{n, v, a}$ & 1.837 & 6 & 520\\
HD 36861 & 195.0520 &  -11.9949 & O8III & 3.3 & 0.120$^\mathrm{s}$ & 0.372 & 6 & 540\\
HD 37128 & 205.2121 &  -17.2416 & B0I & 1.70 & 0.081$^\mathrm{s}$ & 0.356 & 6 & 550\\
HD 32018 & 176.4740 & -9.4588 & B2IV & 7.52 & 0.80$^\mathrm{n, s}$ & 0.429 & 6 & 520\\
HD 16429 & 135.6779 &  1.1453 & O9.5I & 7.85 & 0.92$^\mathrm{a}$ & 1.094 & 6 & 530\\
HD 25940 & 153.6543 & -3.0450 & B3V & 4.00 & 0.173$^\mathrm{s}$ & 0.096 & 6 & 540\\
HD 30614 & 180.3423 &  -14.3712 & O9.5I & 4.30 & 0.315$^\mathrm{a, s, v}$ & 0.813 & 6 & 540\\
HD 43818 & 188.8364 &  4.4803 & B0II & 6.92 & 0.59$^\mathrm{v, a}$ & 1.312 & 6 & 510\\
\hline
\end{tabular}
\label{tab:list}
\end{table*}

\begin{table*}
\centering
\caption{Comparison of the measured equivalent widths of 16 DIB with the data found in the literature. Values in the brackets were excluded from figure \ref{fig:match} as explained in section \ref{sec:comp}.}
\begin{tabular}{c c c c c c c c c c}
\hline\hline
 & \multicolumn{2}{c}{HD 218376} & \multicolumn{3}{c}{HD 30614} & \multicolumn{2}{c}{HD 24398} & \multicolumn{2}{c}{HD 176853}\\
 \cmidrule(lr){2-3}\cmidrule(lr){4-6}\cmidrule(lr){7-8}\cmidrule(lr){9-10}
DIB & this paper & $a$ & this paper & $a$ & $b$ & this paper & $a$ & this paper & $c$\\\hline
4762 & 28.0$\pm$6.8& 13$\pm$2       & 17.6$\pm$3.8 & 17$\pm$3 & 19       & 32.3$\pm$1.8 & 26$\pm$8   & &\\
4964 & 4.1$\pm$1.6 & 6.3$\pm$0.6    & 10.2$\pm$3.1 & 3.5$\pm$0.8 & 2    & 9.0 $\pm$0.9 & 8.8$\pm$1   & 17.7$\pm$3.0 & 17.1$\pm$4.8\\
5512 & 11.4$\pm$5.6 & 5.2$\pm$1     & 2.1$\pm$1.5 & $<$2 &             & 6.6 $\pm$ 2.6 & 4.9$\pm$1   & &\\
5545 & 4.8$\pm$1.8 & 4.4$\pm$1      & 6.9$\pm$2.1 & 9.4$\pm$1 & 8       & 8.5$\pm$6.9 & 6.8$\pm$1.2  & &\\
5705 & 41.7$\pm$7.2 & 45$\pm$5      & 15.8$\pm$5.0 & (54$\pm$10) & 14      & 29.1$\pm$5.9 & 36$\pm$10  & &\\
5780 & 143.5$\pm$5.0 & 146$\pm$8    & 113$\pm$5.3 & 133$\pm$5 & 113      & 112.8$\pm$4.3 & 114$\pm$7 & (205.2$\pm$8.1) & (164.6$\pm$34.8)\\
5797 & 73.1$\pm$4.1 & 61.7$\pm$6    & 44.1$\pm$3.1 & 56$\pm$3 & 34       & 65.3$\pm$2.3 & 77$\pm$5   & 79.9$\pm$4.1 & 94.8$\pm$11.0\\
5850 & & & 19.6$\pm$4.4 && 19                                                  &&                    & 44.7$\pm$5.7 & 53.4$\pm$2.6\\
6090 & & & 7.3$\pm$2.1 && 8                                                            &&            & 14.4$\pm$2.5 & 8.65$\pm$1.8\\
6196 & 15.5$\pm$0.94 & 14.9$\pm$1   & 14.9$\pm$1.4 & 17.2$\pm$1.5 & 15   & 13.2$\pm$2.3 & 16.2$\pm$1 & 13.8$\pm$0.8 & 25.1$\pm$3.4\\
6202 & (39.4$\pm$3.8) & (55$\pm$6)      & (32.5$\pm$3.2) & (63.7$\pm$6) & (21)     & (25.3$\pm$3.59 & (38$\pm$7)   & 39.4$\pm$7.1 & 44.9$\pm$17.8\\
6270 & 20.8$\pm$3.9 & 33.6$\pm$6    & 20.6$\pm$3.9 & 49$\pm$6 & 43       & 9.0$\pm$2.6 & $<$20       & 42.3$\pm$3.8 & 43.9$\pm$12.6\\
6379 & 24.9$\pm$1.8 & 40.5$\pm$2    & 29.2$\pm$2.0 & 34.5$\pm$1 & 26     & 65.2$\pm$3.6 & 67$\pm$3   & 55.9$\pm$3.1 & 59.4$\pm$2.6\\
6445 & & & 12.0$\pm$5.2 && 9                                                 &&                      & 11.6$\pm$4.9 & 13.9$\pm$2.2\\
6613 & 69.9$\pm$3.9 & 68.3$\pm$3    & 68.8$\pm$3.9 & 71.5$\pm$5 & 64     & 59.6$\pm$0.4 & 66$\pm$5   & 95.5$\pm$4.1 & 104.3$\pm$10.5\\
6660 & & & 15.7$\pm$0.3 && 14                                                   &&                   & 16.0$\pm$1.9 & 15.7$\pm$11.7\\\hline
\multicolumn{10}{l}{$a:$ \citet{thorburn03}, $b:$ JD94, $c:$ \citet{lucky13}}
\end{tabular}
\label{tab:match}
\end{table*}

\begin{table*}[!p]
\centering
\caption{Table shows correlation coefficients for each pair of measured DIBs, observed atomic and molecular lines and reddening. Reddening is E(B-V) given in magnitudes. Numbers represent correlation coefficient between equivalent widths of 2 spectral lines or one equivalent width and reddening. For Na and Ca lines a column density was used instead of the equivalent width. The table is split into two parts.}
\begin{tabular}{ccccccccccccccc}
\hline\hline
Line&Redd&4428&4726&4762&4964&5512&5545&5705&5780&5797&5850&6090&6196&6202\\\hline
Redd&1&0.41&0.51&0.74&0.73&0.57&0.64&0.78&0.89&0.84&0.78&0.54&0.90&0.85\\
4428&0.41&1&0.12&0.21&0.22&0.14&0.24&0.35&0.59&0.49&0.41&0.35&0.59&0.53\\
4726&0.51&0.12&1&0.57&0.62&0.60&0.76&0.55&0.47&0.56&0.56&0.28&0.44&0.50\\
4762&0.74&0.21&0.57&1&0.74&0.59&0.47&0.67&0.72&0.69&0.70&0.49&0.74&0.73\\
4964&0.73&0.22&0.62&0.74&1&0.66&0.66&0.59&0.67&0.88&0.90&0.61&0.65&0.73\\
5512&0.57&0.14&0.60&0.59&0.66&1&0.65&0.77&0.45&0.73&0.78&0.60&0.43&0.55\\
5545&0.64&0.24&0.76&0.47&0.66&0.65&1&0.47&0.54&0.72&0.65&0.53&0.55&0.58\\
5705&0.78&0.35&0.55&0.67&0.59&0.77&0.47&1&0.92&0.74&0.72&0.47&0.86&0.87\\
5780&0.89&0.59&0.47&0.72&0.67&0.45&0.54&0.92&1&0.90&0.80&0.63&0.97&0.95\\
5797&0.84&0.49&0.56&0.69&0.88&0.73&0.72&0.74&0.90&1&0.94&0.74&0.91&0.89\\
5850&0.78&0.41&0.56&0.70&0.90&0.78&0.65&0.72&0.80&0.94&1&0.75&0.82&0.83\\
6090&0.54&0.35&0.28&0.49&0.61&0.60&0.53&0.47&0.63&0.74&0.75&1&0.59&0.57\\
6196&0.90&0.59&0.44&0.74&0.65&0.43&0.55&0.86&0.97&0.91&0.82&0.59&1&0.94\\
6202&0.85&0.53&0.50&0.73&0.73&0.55&0.58&0.87&0.95&0.89&0.83&0.57&0.94&1\\
6270&0.87&0.57&0.33&0.63&0.63&0.49&0.53&0.78&0.92&0.90&0.80&0.66&0.92&0.86\\
6379&0.69&0.31&0.52&0.61&0.79&0.59&0.63&0.63&0.72&0.88&0.83&0.85&0.76&0.73\\
6445&0.75&0.51&0.50&0.55&0.64&0.59&0.67&0.54&0.80&0.84&0.72&0.65&0.82&0.73\\
6613&0.84&0.58&0.42&0.60&0.67&0.57&0.60&0.77&0.94&0.93&0.82&0.71&0.96&0.89\\
6660&0.57&0.32&0.34&0.51&0.72&0.57&0.43&0.55&0.61&0.78&0.73&0.56&0.67&0.71\\
7079&0.34&0.11&0.31&0.35&0.21&0.39&0.26&0.40&0.37&0.32&0.34&0.11&0.32&0.48\\
5891 Na {\sc i}&0.84&0.38&0.38&0.60&0.47&0.27&0.48&0.79&0.80&0.64&0.57&0.28&0.86&0.83\\
5897 Na {\sc i}&0.78&0.48&0.53&0.63&0.48&0.49&0.53&0.84&0.91&0.72&0.63&0.37&0.92&0.84\\
4227 Ca {\sc i}&0.35&0.39&0.19&0.27&0.07&0.26&0.25&0.39&0.42&0.35&0.36&0.25&0.45&0.49\\
3934 Ca {\sc ii}&0.62&0.24&0.18&0.49&0.30&0.06&0.38&0.66&0.63&0.42&0.34&0.23&0.62&0.57\\
3969 Ca {\sc ii}&0.75&0.37&0.33&0.61&0.58&0.22&0.47&0.64&0.77&0.67&0.65&0.34&0.79&0.73\\
4300 CH&0.76&0.13&0.39&0.58&0.72&0.32&0.59&0.44&0.63&0.87&0.8&0.59&0.69&0.59\\
4232 CH$^+$&0.70&0.34&0.38&0.60&0.53&0.53&0.49&0.76&0.72&0.70&0.70&0.42&0.72&0.72\\
3874 CN&0.47&0.27&0.52&0.31&0.27&0.59&0.48&0.36&0.33&0.52&0.50&0.26&0.36&0.24\\\hline
\\
\hline\hline
Line&6270&6379&6445&6613&6660&7079&\begin{tabular}{c}5891\\Na {\sc i}\end{tabular}&\begin{tabular}{c}5897\\Na {\sc i}\end{tabular}&\begin{tabular}{c}4227\\Ca {\sc i}\end{tabular}&\begin{tabular}{c}3934\\Ca {\sc ii}\end{tabular}&\begin{tabular}{c}3969\\Ca {\sc ii}\end{tabular}&\begin{tabular}{c}4300\\CH\end{tabular}&\begin{tabular}{c}4232\\CH$^+$\end{tabular}&\begin{tabular}{c}3874\\CN\end{tabular}\\\hline
Redd&0.87&0.69&0.75&0.84&0.57&0.34&0.84&0.78&0.35&0.62&0.75&0.76&0.70&0.47\\
4428&0.57&0.31&0.51&0.58&0.32&0.11&0.38&0.48&0.39&0.24&0.37&0.13&0.34&0.27\\
4726&0.33&0.52&0.50&0.42&0.34&0.31&0.38&0.53&0.19&0.18&0.33&0.39&0.38&0.52\\
4762&0.63&0.61&0.55&0.60&0.51&0.35&0.60&0.63&0.27&0.49&0.61&0.58&0.60&0.31\\
4964&0.63&0.79&0.64&0.67&0.72&0.21&0.47&0.48&0.07&0.30&0.58&0.72&0.53&0.27\\
5512&0.49&0.59&0.59&0.57&0.57&0.39&0.27&0.49&0.26&0.06&0.22&0.32&0.53&0.59\\
5545&0.53&0.63&0.67&0.60&0.43&0.26&0.48&0.53&0.25&0.38&0.47&0.59&0.49&0.48\\
5705&0.78&0.63&0.54&0.77&0.55&0.40&0.79&0.84&0.39&0.66&0.64&0.44&0.76&0.36\\
5780&0.92&0.72&0.80&0.94&0.61&0.37&0.80&0.91&0.42&0.63&0.77&0.63&0.72&0.33\\
5797&0.90&0.88&0.84&0.93&0.78&0.32&0.64&0.72&0.35&0.42&0.67&0.87&0.70&0.52\\
5850&0.80&0.83&0.72&0.82&0.73&0.34&0.57&0.63&0.36&0.34&0.65&0.80&0.70&0.50\\
6090&0.66&0.85&0.65&0.71&0.56&0.11&0.28&0.37&0.25&0.23&0.34&0.59&0.42&0.26\\
6196&0.92&0.76&0.82&0.96&0.67&0.32&0.86&0.92&0.45&0.62&0.79&0.69&0.72&0.36\\
6202&0.86&0.73&0.73&0.89&0.71&0.48&0.83&0.84&0.49&0.57&0.73&0.59&0.72&0.24\\
6270&1&0.69&0.79&0.93&0.68&0.24&0.71&0.76&0.46&0.53&0.69&0.71&0.62&0.31\\
6379&0.69&1&0.76&0.82&0.74&0.31&0.41&0.45&0.36&0.35&0.36&0.79&0.63&0.35\\
6445&0.79&0.76&1&0.86&0.69&0.09&0.56&0.66&0.35&0.50&0.68&0.72&0.38&0.47\\
6613&0.93&0.82&0.86&1&0.83&0.32&0.65&0.73&0.51&0.45&0.66&0.75&0.63&0.25\\
6660&0.68&0.74&0.69&0.83&1&0.28&0.37&0.41&0.47&0.25&0.41&0.76&0.37&0.24\\
7079&0.24&0.31&0.09&0.32&0.28&1&0.44&0.35&0.83&0.07&0.19&-0.03&0.28&-0.06\\
5891 Na {\sc i}&0.71&0.41&0.56&0.65&0.37&0.44&1&0.97&0.28&0.83&0.89&0.44&0.49&0.36\\
5897 Na {\sc i}&0.76&0.45&0.66&0.73&0.41&0.35&0.97&1&0.41&0.77&0.81&0.50&0.58&0.47\\
4227 Ca {\sc i}&0.46&0.36&0.35&0.51&0.47&0.83&0.28&0.41&1&0.61&0.23&0.14&0.31&0.41\\
3934 Ca {\sc ii}&0.53&0.35&0.50&0.45&0.25&0.07&0.83&0.77&0.61&1&0.95&0.40&0.66&0.39\\
3969 Ca {\sc ii}&0.69&0.36&0.68&0.66&0.41&0.19&0.89&0.81&0.23&0.95&1&0.53&0.50&0.18\\
4300 CH&0.71&0.79&0.72&0.75&0.76&0.03&0.44&0.50&0.14&0.40&0.53&1&0.55&0.62\\
4232 CH$^+$&0.62&0.63&0.38&0.63&0.37&0.28&0.49&0.58&0.31&0.66&0.50&0.55&1&0.33\\
3874 CN&0.31&0.35&0.47&0.25&0.24&0.06&0.36&0.47&0.41&0.39&0.18&0.62&0.33&1\\\hline
\end{tabular}
\label{tab:corr}
\end{table*}

\begin{table*}[!p]
\centering
\caption{Table shows probability factors that two lines share the same carrier. For Na and Ca lines a column density was used instead of the equivalent width when calculating the probability factors. Table is split into two parts.}
\begin{tabular}{ccccccccccccccc}
\hline\hline
Line&4428&4726&4762&4964&5512&5545&5705&5780&5797&5850&6090&6196&6202&6270\\\hline
4428&-1&0&-0.01&-0.01&0&0&0.02&0.09&0.06&0.02&0.03&0.08&0.05&0.08\\
4726&0&-1&0.01&0.08&0.23&0.2&0.11&0.01&0.08&0.09&0&-0.01&0.04&-0.03\\
4762&-0.01&0.01&-1&0.21&0.14&0.02&0.05&0.02&0.06&0.09&0.05&0.03&0.07&0\\
4964&-0.01&0.08&0.21&-1&0.14&0.06&0.03&0.03&0.23&0.3&0.16&0.04&0.09&-0.01\\
5512&0&0.23&0.14&0.14&-1&0.17&0.26&0.02&0.15&0.27&0.22&0.01&0.09&0.05\\
5545&0&0.2&0.02&0.06&0.17&-1&0.02&0.01&0.12&0.09&0.11&0&0.02&0.03\\
5705&0.02&0.11&0.05&0.03&0.26&0.02&-1&0.2&0.12&0.14&0.04&0.12&0.14&0.07\\
5780&0.09&0.01&0.02&0.03&0.02&0.01&0.2&-1&0.12&0.1&0.05&0.17&0.15&0.15\\
5797&0.06&0.08&0.06&0.23&0.15&0.12&0.12&0.12&-1&0.29&0.21&0.14&0.13&0.13\\
5850&0.02&0.09&0.09&0.3&0.27&0.09&0.14&0.1&0.29&-1&0.25&0.11&0.13&0.09\\
6090&0.03&0&0.05&0.16&0.22&0.11&0.04&0.05&0.21&0.25&-1&0.06&0.01&0.14\\
6196&0.08&-0.01&0.03&0.04&0.01&0&0.12&0.17&0.14&0.11&0.06&-1&0.13&0.15\\
6202&0.05&0.04&0.07&0.09&0.09&0.02&0.14&0.15&0.13&0.13&0.01&0.13&-1&0.07\\
6270&0.08&-0.03&0&-0.01&0.05&0.03&0.07&0.15&0.13&0.09&0.14&0.15&0.07&-1\\
6379&-0.01&0.02&0.07&0.21&0.15&0.14&0.05&0.04&0.26&0.23&0.28&0.07&0.06&0.08\\
6445&0.1&0.02&0.02&0.08&0.09&0.18&0.02&0.04&0.15&0.09&0.19&0.08&0.05&0.14\\
6613&0.11&0&-0.01&0.08&0.04&0.06&0.08&0.13&0.18&0.15&0.18&0.15&0.1&0.2\\
6660&0.03&0.02&0.05&0.22&0.18&0.04&0.08&0.06&0.23&0.22&0.15&0.11&0.08&0.07\\
7079&0.01&0.03&0&0&0.01&0.02&0&0&0.01&0.01&0&0&0.01&0\\
5891 Na {\sc i}&0.01&0&0.03&-0.09&-0.02&0.01&0.1&0.1&0.01&-0.01&-0.04&0.08&0.11&0.05\\
5897 Na {\sc i}&0.05&0.07&0&-0.02&0.01&0.04&0.16&0.13&0.05&0.01&-0.02&0.1&0.09&0.09\\
4227 Ca {\sc i}&0.04&0&-0.01&0&0&0.02&0.01&-0.01&-0.01&-0.01&0&0&-0.01&-0.01\\
3934 Ca {\sc ii}&0&-0.02&0.03&-0.02&-0.01&0.02&0.03&0.03&-0.03&-0.03&-0.02&0.03&0&0\\
3969 Ca {\sc ii}&0.02&-0.02&0.1&0.06&-0.02&0.05&0.09&0.07&0.02&0.02&0&0.06&0.07&0.04\\
4300 CH&-0.01&0&0.03&0.13&-0.01&0.06&-0.03&-0.02&0.13&0.09&0.08&0.02&-0.06&0.04\\
4232 CH$^+$&0.01&0.02&0.04&0.06&0.13&0.04&0.19&0.1&0.06&0.11&0.03&0.11&0.07&0.02\\
3874 CN&0.04&0.15&-0.02&-0.01&0.09&0.08&0.01&-0.02&0.01&-0.01&0.01&-0.02&-0.02&-0.02\\\hline
\\
\hline\hline
Line&6379&6445&6613&6660&7079&\begin{tabular}{c}5891\\Na {\sc i}\end{tabular}&\begin{tabular}{c}5897\\Na {\sc i}\end{tabular}&\begin{tabular}{c}4227\\Ca {\sc i}\end{tabular}&\begin{tabular}{c}3934\\Ca {\sc ii}\end{tabular}&\begin{tabular}{c}3969\\Ca {\sc ii}\end{tabular}&\begin{tabular}{c}4300\\CH\end{tabular}&\begin{tabular}{c}4232\\CH$^+$\end{tabular}&\begin{tabular}{c}3874\\CN\end{tabular}\\\hline
4428&-0.01&0.1&0.11&0.03&0.01&0.01&0.05&0.04&0&0.02&-0.01&0.01&0.04&\\
4726&0.02&0.02&0&0.02&0.03&0&0.07&0&-0.02&-0.02&0&0.02&0.15&\\
4762&0.07&0.02&-0.01&0.05&0&0.03&0.01&-0.01&0.03&0.1&0.03&0.04&-0.02&\\
4964&0.21&0.08&0.08&0.22&0&-0.09&-0.02&0&-0.02&0.06&0.13&0.06&-0.01&\\
5512&0.15&0.09&0.04&0.18&0.01&-0.02&0.01&0&-0.01&-0.02&-0.01&0.13&0.09&\\
5545&0.14&0.18&0.06&0.04&0.02&0.01&0.04&0.02&0.02&0.05&0.06&0.04&0.08&\\
5705&0.05&0.02&0.08&0.08&0&0.1&0.16&0.01&0.03&0.09&-0.03&0.19&0.01&\\
5780&0.04&0.04&0.13&0.06&0&0.1&0.14&-0.01&0.03&0.07&-0.02&0.1&-0.02&\\
5797&0.26&0.15&0.18&0.23&0.01&0.01&0.05&-0.01&-0.03&0.02&0.13&0.06&0.01&\\
5850&0.23&0.09&0.15&0.22&0.01&-0.01&0.02&-0.01&-0.03&0.02&0.09&0.11&-0.01&\\
6090&0.28&0.19&0.18&0.15&0&-0.04&-0.01&0&-0.02&0&0.08&0.03&0.01&\\
6196&0.07&0.08&0.15&0.11&0&0.08&0.11&0&0.03&0.06&0.02&0.11&-0.02&\\
6202&0.06&0.05&0.1&0.08&0.01&0.11&0.1&-0.01&0&0.07&-0.06&0.07&-0.02&\\
6270&0.08&0.14&0.2&0.07&0&0.05&0.09&-0.01&0&0.04&0.04&0.02&-0.02&\\
6379&-1&0.17&0.18&0.26&0.04&-0.04&-0.03&0.04&-0.02&-0.02&0.14&0.07&0.03&\\
6445&0.17&-1&0.19&0.15&0.01&0&0.06&0&0.04&0.09&0.03&-0.03&0.01&\\
6613&0.18&0.19&-1&0.22&0&0.01&0.06&-0.01&-0.02&0.02&0.06&0.06&-0.02&\\
6660&0.26&0.15&0.22&-1&0&-0.02&-0.01&0.01&-0.03&0&0.11&0&0&\\
7079&0.04&0.01&0&0&-1&0&0&0.07&0.01&0&0&0.02&0.01&\\
5891 Na {\sc i}&-0.04&0&0.01&-0.02&0&-1&0.36&0.01&0.14&0.09&-0.06&0&-0.01&\\
5897 Na {\sc i}&-0.04&0.06&0.06&-0.01&0&0.36&-1&0&0.12&0.15&-0.04&0.02&0.04&\\
4227 Ca {\sc i}&0.04&0&-0.01&0.01&0.07&0.01&0&-1&0.29&0.02&-0.01&0.05&0.06&\\
3934 Ca {\sc ii}&-0.02&0.04&-0.02&-0.03&0.01&0.14&0.12&0.29&-1&0.38&0&0.05&0.08&\\
3969 Ca {\sc ii}&-0.02&0.09&0.02&0&0&0.09&0.15&0.02&0.38&-1&-0.01&0.05&-0.01&\\
4300 CH&0.14&0.03&0.06&0.11&0&-0.06&-0.04&-0.01&0&-0.01&-1&-0.02&0.07&\\
4232 CH$^+$&0.07&-0.03&0.06&0&0.02&0&0.02&0.05&0.05&0.05&-0.02&-1&-0.01&\\
3874 CN&0.03&0.01&-0.02&0&0.01&-0.01&0.03&0.06&0.08&-0.01&0.07&-0.01&-1&\\\hline
\end{tabular}
\label{tab:prob}
\end{table*}

\begin{table*}[!p]
\centering
\caption{Table shows linear relations between reddening and equivalent width of each DIB for $\sigma$ and $\zeta$ sightlines. Equation is $W_{eq}=a\cdot \mathrm{E(B-V)}+b$ if reddening is in magnitudes and $W_{eqw}$ is in \AA . One sigma errors are given for $a$ and $b$, joined by the values of the correlation coefficient and reduced $\chi^2$.}
\begin{tabular}{cccccccc}
\hline\hline
DIB & Sightline& $a$ & $b$ & $a$ Error & $b$ Error & Corr & Reduced $\chi^2$\\\hline
\multirow{2}{*}{4762} &$\zeta$ &0.1247 &-0.0158 &0.0207 &0.0131 &0.79 &3.23\\
 &$\sigma$ &0.0755 &0.0029 &0.0103 &0.0069 &0.83 &6.55\\[0.2cm]
\multirow{2}{*}{4964} &$\zeta$ &0.0419 &-0.0045 &0.0068 &0.0043 &0.84 &7.97\\
 &$\sigma$ &0.0169 &0.0023 &0.003 &0.002 &0.76 &6.50\\[0.2cm]
\multirow{2}{*}{5512} &$\zeta$ &0.0242 &-0.0034 &0.0054 &0.0033 &0.76 &9.92\\
 &$\sigma$ &0.0082 &0.003 &0.0036 &0.0028 &0.52 &12.9\\[0.2cm]
\multirow{2}{*}{5545} &$\zeta$ &0.0495 &0.0024 &0.0112 &0.0068 &0.63 &9.22\\
 &$\sigma$ &0.0328 &0.0033 &0.0063 &0.0041 &0.62 &10.09\\[0.2cm]
\multirow{2}{*}{5705} &$\zeta$ &0.12 &-0.0104 &0.0312 &0.0191 &0.77 &17.19\\
 &$\sigma$ &0.1098 &0.0061 &0.0174 &0.0114 &0.74 &9.6\\[0.2cm]
\multirow{2}{*}{5780} &$\zeta$ &0.6029 &-0.0834 &0.0568 &0.0344 &0.92 &4.99\\
 &$\sigma$ &0.5058 &-0.0142 &0.0503 &0.0315 &0.87 &13.61\\[0.2cm]
\multirow{2}{*}{5797} &$\zeta$ &0.1992 &-0.0182 &0.0202 &0.0124 &0.91 &5.64\\
 &$\sigma$ &0.1239 &-0.0029 &0.0139 &0.0089 &0.85 &16.81\\[0.2cm]
\multirow{2}{*}{5850} &$\zeta$ &0.0898 &-0.006 &0.015 &0.0091 &0.81 &10.69\\
 &$\sigma$ &0.0502 &0.0014 &0.0054 &0.0035 &0.87 &5.3\\[0.2cm]
\multirow{2}{*}{6090} &$\zeta$ &0.0266 &0.0015 &0.0042 &0.0026 &0.78 &7.43\\
 &$\sigma$ &0.0129 &0.0041 &0.0039 &0.0028 &0.51 &14.22\\[0.2cm]
\multirow{2}{*}{6196} &$\zeta$ &0.0603 &-0.0062 &0.0068 &0.0042 &0.9 &6.83\\
 &$\sigma$ &0.0479 &0.0009 &0.004 &0.0026 &0.91 &6.76\\[0.2cm]
\multirow{2}{*}{6202} &$\zeta$ &0.0955 &-0.0054 &0.0142 &0.0082 &0.85 &6.47\\
 &$\sigma$ &0.0907 &0.0032 &0.0104 &0.0066 &0.86 &7.65\\[0.2cm]
\multirow{2}{*}{6270} &$\zeta$ &0.1158 &-0.0141 &0.0131 &0.0082 &0.9 &6.24\\
 &$\sigma$ &0.0844 &-0.0006 &0.0124 &0.0082 &0.81 &13.28\\[0.2cm]
\multirow{2}{*}{6379} &$\zeta$ &0.1255 &-0.0093 &0.0207 &0.0126 &0.78 &13.61\\
 &$\sigma$ &0.0572 &0.0043 &0.0118 &0.0078 &0.66 &40.18\\[0.2cm]
\multirow{2}{*}{6445} &$\zeta$ &0.0356 &-0.0029 &0.0067 &0.0041 &0.75 &4.72\\
 &$\sigma$ &0.023 &0.0014 &0.0038 &0.0025 &0.64 &13.28\\[0.2cm]
\multirow{2}{*}{6613} &$\zeta$ &0.2576 &-0.0171 &0.0243 &0.0148 &0.92 &5.26\\
 &$\sigma$ &0.1917 &0.0031 &0.0256 &0.0165 &0.79 &28.45\\[0.2cm]
\multirow{2}{*}{6660} &$\zeta$ &0.0485 &-0.0036 &0.0056 &0.0034 &0.9 &5.32\\
 &$\sigma$ &0.0135 &0.0087 &0.0042 &0.0028 &0.52 &34.79\\\hline
\end{tabular}
\label{tab:rel}
\end{table*}

\begin{table*}[!p]
\centering
\caption{Example of data and format of electronically provided parameters. Qantities are defined in section \ref{sec:asim}.}
\begin{tabular}{ccccccccc}
\hline\hline
\multicolumn{9}{c}{HD210839}\\\hline
DIB & Position & \begin{tabular}{c}Position\\Error\end{tabular} & W$_{eq}$ & \begin{tabular}{c} W$_{eq}$\\Error\end{tabular} & Width & \begin{tabular}{c}Width\\Error\end{tabular} & Asymmetry & \begin{tabular}{c}Asymmetry\\Error\end{tabular}\\\hline
 & \AA & \AA & m\AA & m\AA & \AA & \AA & & \\\hline
4428 & 4427.16 & 0.08 & 202.4 & 24.3 & 5.671 & 0.127 & 0.26 & 0.25\\
4726 & 4725.97 & 0.03 & 56.0 & 7.6 & 2.069 & 0.027 & -0.59 & 0.05\\
4762 & 4762.61 & 0.07 & 64.9 & 19.1 & 3.657 & 0.308 & 0.48 & 0.62\\
4964 & 4963.61 & 0.02 & 6.0 & 1.5 & 0.414 & 0.016 & -0.26 & 0.03\\
5512 & 5512.4 & 0.07 & 1.6 & 1.9 & 0.534 & 0.088 & 0.0 & 0.18\\
5545 & 5544.75 & 0.02 & 12.8 & 2.7 & 0.752 & 0.02 & -0.31 & 0.04\\
5705 & 5704.55 & 0.04 & 51.5 & 8.8 & 2.327 & 0.049 & -0.34 & 0.1\\
5780 & 5779.99 & 0.01 & 264.5 & 7.1 & 1.909 & 0.009 & -0.19 & 0.02\\
5797 & 5796.71 & 0.01 & 78.3 & 3.5 & 0.896 & 0.007 & -0.26 & 0.01\\
5850 & 5849.56 & 0.02 & 25.4 & 3.2 & 0.911 & 0.016 & 0.07 & 0.03\\
6090 & 6089.54 & 0.02 & 12.2 & 2.1 & 0.566 & 0.02 & 0.03 & 0.04\\
6196 & 6195.65 & 0.01 & 37.6 & 1.8 & 0.613 & 0.031 & -0.32 & 0.04\\
6202 & 6202.85 & 0.01 & 53.9 & 3.4 & 1.129 & 0.04 & 0.18 & 0.09\\
6270 & 6269.55 & 0.03 & 71.0 & 6.7 & 1.644 & 0.037 & 0.22 & 0.07\\
6379 & 6378.99 & 0.01 & 48.4 & 2.1 & 0.563 & 0.005 & -0.02 & 0.01\\
6445 & 6445.01 & 0.02 & 17.3 & 2.5 & 0.653 & 0.022 & 0.21 & 0.04\\
6613 & 6613.1 & 0.01 & 163.2 & 3.0 & 1.008 & 0.053 & -0.25 & 0.11\\
6660 & 6660.47 & 0.01 & 27.1 & 2.8 & 0.757 & 0.015 & 0.37 & 0.03\\
7079 & 7077.94 & 0.03 & 116.2 & 9.6 & 1.986 & 0.026 & -0.74 & 0.05\\
5891 Na {\sc i} & 5889.64 & 0.01 & 688.6 & 8.3 & 0.626 & 0.003 & 0.0 & 0.0\\
5897 Na {\sc i} & 5895.61 & 0.01 & 622.1 & 22.2 & 0.606 & 0.008 & 0.0 & 0.0\\
4227 Ca {\sc i} & 4226.63 & 0.01 & 9.7 & 1.3 & 0.298 & 0.006 & 0.0 & 0.0\\
3934 Ca {\sc ii} & 3933.52 & 0.02 & 232.5 & 3.7 & 0.393 & 0.002 & 0.0 & 0.0\\
3969 Ca {\sc ii} & 3968.33 & 0.01 & 114.3 & 1.3 & 0.45 & 0.08 & 0.0 & 0.0\\
4300 CH & 4300.14 & 0.02 & 18.0 & 1.1 & 0.254 & 0.002 & 0.0 & 0.0\\
4232 CH$^+$& 4232.39 & 0.02 & 11.0 & 1.2 & 0.305 & 0.004 & 0.0 & 0.0\\
3874 CN& 3874.38 & 0.01 & 9.1 & 1.8 & 0.292 & 0.014 & 0.0 & 0.0\\\hline
\end{tabular}
\label{tab:example}
\end{table*}

\end{document}